\documentclass[11pt]{article}

\usepackage{amsmath}
\usepackage{graphicx}
\usepackage{amsfonts}
\usepackage{amssymb}
\usepackage{epsfig}
\usepackage{color}
\usepackage{psfrag}

 \newtheorem{theorem}{Theorem}[section]
    \newtheorem{rema}{Remark}[section]
    \newtheorem{assump}{Assumptions}[section]
    \newtheorem{propo}[rema]{Proposition}

\newcommand{\EQ}{\begin{equation}}
\newcommand{\EN}{\end{equation}}

\newcommand{\bc}{\begin{center}}
\newcommand{\ec}{\end{center}}
\def\ba#1{\begin{array}{#1}\displaystyle}
\newcommand{\ea}{\end{array}}

\newcommand{\beq}{\begin{equation}}
\newcommand{\eeq}{\end{equation}}
\newcommand{\beqa}{\begin{eqnarray}}
\newcommand{\eeqa}{\end{eqnarray}}
\newcommand{\no}{\nonumber}
\newcommand{\n}{\nonumber\\}
\newcommand{\bi}{\begin{itemize}}
\newcommand{\ei}{\end{itemize}}
\def\lt#1{\left#1}
\def\rt#1{\right#1}
\def\t#1{\tilde{#1}}
\def\h#1{\hat{#1}}
\def\b#1{\bar{#1}}
\def\frc#1#2{\frac{#1}{#2}}

\newcommand{\p}{\partial}

\newcommand{\bra}{\langle}
\newcommand{\ket}{\rangle}
\newcommand{\Z}{{\mathbb{Z}}}
\newcommand{\N}{{\mathbb{N}}}
\newcommand{\R}{{\mathbb{R}}}
\newcommand{\C}{{\mathbb{C}}}
\newcommand{\uH}{{\mathbb{H}}}
\newcommand{\uD}{{\mathbb{D}}}

\newcommand{\Or}{{\cal O}}

\newcommand{\ep}{\epsilon}

\newcommand{\bs}{\setminus}

\newcommand{\id}{{\rm id}}

\newcommand{\halmos}{\rule{1ex}{1.4ex}}
\newcommand{\eproof}{\hspace*{\fill}\mbox{$\halmos$}}

\begin{document}

\setcounter{page}{0} \topmargin 0pt
\renewcommand{\thefootnote}{\arabic{footnote}}
\newpage
\setcounter{page}{0}

\begin{titlepage}

\vspace{0.5cm}
\begin{center}
{\Large {\bf Identification of the stress-energy tensor through}}\\
{\Large {\bf conformal restriction in SLE and related processes}}

\vspace{2cm}
{\large B. Doyon$^{a}$, V. Riva$^{a,b}$ and J. Cardy$^{a,c}$} \\
\vspace{0.5cm} {\em $^{a}$Rudolf Peierls Centre for Theoretical
Physics, 1 Keble Road, Oxford OX1 3NP, UK} \\\vspace{0.3cm} {\em
$^{b}$Wolfson College, Oxford} \\\vspace{0.3cm} {\em $^{c}$All
Souls College, Oxford}

\end{center}

\vspace{1cm}

\begin{abstract}
We derive the Ward identities of Conformal Field Theory (CFT)
within the framework of Schramm-Loewner Evolution (SLE) and some
related processes. This result, inspired by the observation that
particular events of SLE have the correct physical spin and
scaling dimension, and proved through the conformal restriction
property, leads to the identification of some probabilities with
correlation functions involving the bulk stress-energy tensor.
Being based on conformal restriction, the derivation holds for SLE
only at the value $\kappa=8/3$, which corresponds to the central
charge $c=0$ and the case when loops are suppressed in the
corresponding $O(n)$ model.

\vspace{2cm}

\hrulefill

E-mail addresses: j.cardy1, b.doyon1, v.riva1@physics.ox.ac.uk

\end{abstract}

\end{titlepage}

\newpage

\section{Introduction}
\label{intro}

The description of two-dimensional statistical models at their
critical points in terms of Conformal Field Theories (CFT) is one
of the most fruitful achievements of theoretical physics
\cite{BPZ} (for a pedagogical account, see \cite{yellow}). In the
past twenty years, remarkable exact results on universal
quantities like critical exponents have been obtained within this
framework. However, some issues have been only partially
understood. Besides the lack of mathematical rigor in relating
statistical models to CFT, the language of CFT is not best suited
to the description of the geometrical aspects of conformal
symmetry, being formulated upon the concept of local operators.
Moreover, a clear and rigorous geometrical definition of local
conformal invariance (and of its breaking by an anomaly,
quantified by the \lq central charge' $c$) is missing in CFT. An
important progress in filling these gaps has recently been
achieved in the context of probability theory and stochastic
analysis, with a new approach to critical phenomena centered on
Schramm-Loewner Evolution (SLE) \cite{loewner,schramm} (for a
review aimed at theoretical physicists, see \cite{review}). In a
nutshell, SLE is a way of constructing measures on random curves
which satisfy the expected properties of domain walls of critical
statistical systems in the continuum limit. It turns out that such
measures form a family described by one real parameter $\kappa$.
Different values of $\kappa$ are expected to correspond to
different statistical systems. The chordal version of SLE, which
is the only one considered in detail in this paper, defines a
measure on curves conditioned to start and end at distinct points
on the boundary of a simply connected domain in $\C$, which can be
conventionally chosen to be the upper half plane $\uH$ by virtue
of conformal invariance.

A natural question which arises is the precise relation between
SLE and CFT. A first step in this direction was made by noticing
\cite{BB} that the Fokker-Planck-type equations obtained from SLE
are closely related to second order differential equations
satisfied by certain CFT correlation functions involving the
so-called \lq boundary condition changing operators' $\phi_{2,1}$
\cite{JC1984}. This implies a precise relation between
probabilities in SLE and correlation functions in CFT with the
boundary operator $\phi_{2,1}$ inserted at the points where the
SLE curve starts and ends. An important consequence of the above
identification is the relation between the parameter $\kappa$ and
the central charge $c$ of CFT:
\begin{equation}\label{c(kappa)}
c\,=\,\frac{(3\kappa-8)(6-\kappa)}{2\kappa}\;.
\end{equation}
However, a deeper insight requires the identification of
correlation functions involving other kinds of operators, inserted
not only at the boundary but especially in the bulk of the domain.
Particularly significant in CFT are holomorphic operators, which
transform non-triviallly under only one of the two copies of the
underlying Virasoro algebra. Among holomorphic operators, the most
important is the stress-energy tensor, the generator of conformal
transformations, whose Ward identities are equivalent to the
statement that scaling operators should be classified according to
highest-weight representations of the Virasoro algebra.

This paper deals with the identification of some probabilities in
the SLE context with CFT correlation functions involving the bulk
stress-energy tensor $T$. More precisely, we consider the joint
probability that the SLE intersects a number of short segments in
the bulk, of lengths $\{\epsilon_j\}$ and centered about points
$\{w_j\}$, at inclinations to some fixed axis characterized by
angular variables $\{\theta_j\}$. One can then investigate the
features of the Fourier components of this probability with
respect to $\theta$, which are labelled by a variable $n$ which
has the properties of conformal \lq\lq spin". We study the leading
behavior of each component as $\epsilon\to0$, which we assume is a
power law. For instance, the leading power of the spin zero
Fourier component is $2-d_f$, where $d_f$ is the fractal dimension
of SLE, rigorously computed in \cite{fracdim}. It is natural to
guess a relation between the spin-2 Fourier component and the
holomorphic stress-energy tensor, which is an operator carrying
spin 2. The central result of the paper is the justification of
this correspondence by proving that the second Fourier components
of the above described probabilities satisfy the so-called
conformal Ward identities, which are the mathematical
formalization of the fact that the stress energy tensor generates
conformal transformations. The instrumental tool in our proof is
the so-called conformal restriction property \cite{LSW}, which
refers not only to SLE, but also to more general random processes
on the plane. Actually, conformal restriction has already been used in
\cite{FW} to derive the Ward identities on the boundary, but we
shall implement it differently using a method which is not
restricted to work on the boundary only. As a by-product of our
analysis, we obtain slightly more general results for the boundary
case itself, with a more accurate interpretation in terms of CFT
correlators.

The paper is organized as follows. In Section\,\ref{sectresult} we
present our assumptions and the main result of the paper. In
Section\,\ref{sectmotivation} we describe the SLE problem under
consideration, analyzing the probability that a curve passes
between the ending points of a segment and the cases in which it
suggests the identification with a holomorphic operator in CFT. We
also discuss the analogies and differences between the events of
passing between the ending points of the segment or intersecting
the segment. Section\,\ref{sectRestr} contains the proof of our
main result, valid for conformal restriction measures. In
Section\,\ref{sectinterpr} we specialize the result to SLE$_{8/3}$
and we interpret it in the CFT language, showing that it
corresponds to the Ward identities at $c=0$.
Section\,\ref{sectbdry} discusses the boundary case and possible
generalizations of our result to CFT with $c< 0$. Finally, in
Section\,\ref{conclusions} we present our conclusions. The paper
also includes four Appendices which present technical results
useful for the general discussion.

\section{Assumptions and main
results}\label{sectresult}\setcounter{equation}{0}

We first state our assumptions. Consider a measure on a random
connected set $K\subset \uH$ with $\{0,\infty\}\subset \b{K}$
satisfying {\em conformal restriction}. Conformal restriction
measures were defined and studied in \cite{LSW}, and their
properties will be summarized in Section \ref{sectRestr}. In
particular, they are characterized by a real number $h$, called
restriction exponent, which will be defined in (\ref{restrexp})
and which will explicitly appear in our main result. Many
properties of conformal restriction measures are known, but we
need to assume some ``smoothness'' properties of probabilities.
Although, to our knowledge, these properties were not fully
assessed yet in the literature, it is our expectation that their
proof, in the case of SLE, is a matter of a technical analysis, and that
for other conformal restriction measures, they are essential for the definition
of a local stress-energy tensor.

More precisely, consider indicator events associated to the set
$K$ depending on points $z_j\in\uH$: events that the set $K$ is to
the right of $z_j$, to the left of $z_j$, or that $z_j$ is inside
$K$ (i.e.~$z_j$ is included in the filling of $K$), and consider
the generic probability
\[
    P(K\subset \uH\bs D_k, z_1,\ldots,z_l)
\]
where $D_k\subset \uH$, $\{0,\infty\}\subset \uH\bs D_k$, $\uH\bs
D_k$ is topologically the upper half plane with $k$ holes and the
boundary $\partial D_k$ is piecewise smooth. Commas represent
intersection of events and $z_1,\ldots,z_l$ represent indicator
events (the probability is not expected to be a holomorphic
function of $z_1,\ldots,z_l$, but for notational convenience we
will not write explicitly its dependence on
$\b{z}_1,\ldots,\b{z}_l$). By conformal restriction, this only
depends on the coordinates in the moduli space of $\uH\bs (D_k\cup
\{z_1,\ldots,z_l\})$ with $0$ and $\infty$ fixed. These
coordinates can be taken as the positions of $l$ points and the
central positions and lengths of $k$ horizontal slits in $\uH$ (up
to an overall scale transformation). Then, our assumptions amount
to a statement of smoothness in moduli space, precisely:
\begin{assump} \label{assump}
With $P(K\subset \uH\bs D_k, z_1,\ldots,z_l)$ and its coordinates
in the moduli space as described above, we have
\begin{enumerate}
\item The singularities of the first derivative of $P(K\subset
\uH\bs D_k, z_1,\ldots,z_l)$ in the moduli space may occur on the
hyper-planes corresponding to the situations where 1) any slit or
any point touches the real axis, 2) any two or more of the slits
or points enter in contact with each other, or 3) the length of
any slit is sent to zero. In any direction at points in the moduli
space away from the singular planes, and on the singular planes
parallel to them, the probability is differentiable at least once.
\item The limit towards a singular plane commutes with the
derivative in any direction parallel to the singular plane at that
point. \item The probability behave, when the length $\ep$ of one
of the slit is sent to 0, as the same probability with this slit
missing (denoting the corresponding domain by $D_{k-1}$) plus a
correction which is a power of the length of the slit:
\[
    P(K\subset \uH\bs D_k, z_1,\ldots,z_l) - P(K\subset \uH\bs D_{k-1}, z_1,\ldots,z_l) = O(\ep^{2-d})
\]
for some $d<2$. (In the case of SLE$_{8/3}$, which is a conformal
restriction measure, $d$ is known to be $4/3$, the fractal
dimension of the curve.)
\end{enumerate}
\end{assump}
These assumptions can easily be verified for the particular case
of the probability $P(z)$ in SLE$_{8/3}$ using the arguments of
Schramm \cite{schramm} (here, $P(z)$ can be the probability that
the SLE curve be to the right of the point $z$, or the probability
that it be to its left). An idea of the general proof (at least
for Point 1) could be as follows. Consider first the case of
SLE$_{8/3}$. This is the unique conformal restriction measure that
is supported on simple curves. It can be constructed by
dynamically growing a curve using a Loewner map with a
time-dependent driving term proportional to a one-dimensional
standard Brownian motion started at 0. This measure has the
property that when it is restricted to the curve having a given
shape $\Gamma$ from 0 to any point inside $\uH$, then it is equal
to the measure obtained by a conformal transformation, through a
Loewner map, from $\uH\bs\Gamma$ to $\uH$. Ito's calculus then
tells us that the derivative, in the moduli space, of the
probabilities considered above in the direction specified by a
small Loewner map $z \mapsto z+ dt/z$ exists: this is at the basis
of the derivation of the ``SLE equation'' for such probabilities.
In fact, Ito's calculus tells us more: for every curve $\Gamma$,
the Loewner maps of all sub-curves starting at 0 define a path in
the moduli space. Then, the derivatives in the moduli space along
all these paths exist. The proof would need to show that taking
all curves $\Gamma$ which can restrict the measure of SLE$_{8/3}$,
one can describe paths such that at any non-singular point in the
moduli space, all directions occur. We expect that
Assumptions\,\ref{assump} also hold for other conformal
restriction measures. In those cases, one needs other explicit
constructions along the lines discussed above. It is worth noting that
Point 3 is probably the most delicate: the set $K$ cannot be space-filling.
We will briefly come back to this in the context of a certain conformal
restriction measure constructed (by adding brownian bubbles to SLE) in \cite{LSW}.

In order to state our results, we need to introduce some objects
and some notations. We first define a family ${\cal E}$ of simply
connected domains in $\uH$ whose members $E_{w,\ep,\theta}[D]$ are
parametrized by $w\in\uH$ (that is, with ${\rm Im}(w)>0$), $\ep>0$
and $\theta\in[0,2\pi]$, as well as a simply connected domain $D$
of a certain type, with $\partial D$ piecewise smooth. More
precisely, the members of ${\cal E}$ are defined by
\beq\label{defE}
    g_{w,\ep,\theta}(\uH\bs S_{w,\ep}(D)) = \uH \bs E_{w,\ep,\theta}[D]
\eeq where $S_{w,\ep}$ is a conformal map that scales by $\ep$
with center at $w$: $S_{w,\ep}(z) = w+\ep(z-w)$ with
$S_{w,\ep}(D)\in \uH$. The conformal transformations
$g_{w,\ep,\theta}$ are defined by
\begin{equation}\label{g}
    g_{w,\ep,\theta}(z)\,=\,z\,+\,\frac{\epsilon^2}{16}\,\frac{e^{2i\theta}}{w-z}\,+\,
    \frac{\epsilon^2}{16}\,\frac{e^{-2i\theta}}{\bar{w}-z}
    \,-\,\frac{\epsilon^2}{16}\,\frac{e^{2i\theta}}{w}\,-\,\frac{\epsilon^2}{16}\,\frac{e^{-2i\theta}}{\bar{w}}
\end{equation}
and we can take any simply connected domain $D\in
S_{w,\ep}^{-1}(\uH)$ such that the right-hand side of Eq.
(\ref{defE}) indeed is a subset of $\uH$. In particular, the
domain $S_{w,\ep}(D)$ must include the branch points of
$g_{w,\ep,\theta}$ that are situated in $\uH$, which means that
$D$ must include the points at the positions $w\pm
\frc{i}4e^{i\theta} + O(\ep)$ for small $\ep$. Note also that for
any $D$ which strictly contains the disk of radius $1/4$ centered
at $w$, there exists an $\ep[D]$ such that for all $0<\ep<\ep[D]$,
$E_{w,\ep,\theta}[D]$ exists. If $D$ is a disk centered at $w$ of
radius $b/4$ for some $b>1$, then the boundary of
$E_{w,\ep,\theta}[D]$ describes an ellipse centered at $w$ of
major axis $\left(b+\frac{1}{b}\right)\frac{\epsilon}{2}$ and
minor axis $\left(b-\frac{1}{b}\right)\frac{\epsilon}{2}$, plus a
deformation of order $\ep^2$ of this ellipse:
\begin{eqnarray}\label{gshape}
    g_{w,\ep,\theta}\left(w+\frac{b\epsilon}{4}\,e^{i\alpha+i\theta}\right) &=&
    w+\frac{\epsilon}{4}e^{i\theta+i\pi/2}\,\left(b+\frac{1}{b}\right)\,\sin\alpha
    +\frac{\epsilon}{4}e^{i\theta}\,\left(b-\frac{1}{b}\right)\,\cos\alpha
    \,+\,\\\nonumber
    &-&\frac{\epsilon^2}{16}\,\frac{e^{2i\theta}}{w}
    \,-\,\frac{\epsilon^2}{16}\,\frac{e^{-2i\theta}}{\bar{w}}\,+\,
        \frac{\epsilon^2}{16}\,\frac{e^{-2i\theta}}{\bar{w}-w}
    \,+\,b\frac{\epsilon^3}{64}\,\frac{e^{i\alpha+i\theta}}{(\bar{w}-w)^2}\,+\,O(\epsilon^4)\;,
\end{eqnarray}
where $\alpha\in[0,2\pi]$. The major axis of the ellipse makes an
angle $\theta$ with respect to the positive imaginary direction.
The ellipse becomes, as $b\to 1$, a segment of length $\ep$
centered at $w$ and of angle $\theta$ with respect to the
imaginary direction.

We will consider the event that the set $K$ intersects a member
$E_{w,\ep,\theta}[D]$ of the family ${\cal E}$ described above. In
fact, since our results will be independent of the exact form of
$D$, we will drop the explicit dependence on $D$. Let us then
denote, for $n_1,\ldots,n_k\in\Z$: \beqa &&
    Q^{(k,l)}_{n_1,\ldots,n_k}(w_1,\ldots,w_k,z_1,\ldots,z_l) = \n && \qquad\qquad
    \lt(\frc{8}\pi\rt)^k\,\lim_{\ep_1,\ldots,\ep_k\to0}
    \ep_1^{-|n_1|}\cdots \ep_k^{-|n_k|} \;
    \int_0^{2\pi} d\theta_1\,e^{-in_1\theta_1} \cdots
    \int_0^{2\pi} d\theta_k\,e^{-in_k\theta_k}\; \cdot \n && \qquad\qquad\qquad\qquad \cdot \;
    P(K\cap E_{w_1,\ep_1,\theta_1}\neq \emptyset,\ldots,K\cap E_{w_k,\ep_k,\theta_k}\neq\emptyset,
        z_1,\ldots,z_l)~, \label{Qmultiple}
\eeqa whenever this limit exists. Although we expect that it does
exist for all $n_1,\ldots,n_k \in\Z\backslash\{0\}$, we will only
need a subset of these (and our main theorem applies only to a
particular case); we introduce this general notation in order to
make contact with the motivations which led to our result. In
particular, we expect that the numbers $n_i$ correspond to the
\lq\lq spin" and that their absolute values $|n_i|$ correspond to
the \lq\lq scaling dimension" of holomorphic (or antiholomorphic)
operators in CFT\footnote{More precisely, in the CFT language, one
associates to primary operators the conformal dimensions (real
numbers) $h$ and $\b{h}$, in terms of which the spin is $h-\b{h}$
and the scaling dimension is $h+\b{h}$. Holomorphic field are
those for which $\b{h}=0$ and anti-holomorphic fields are those
for which $h=0$.}. Hence, we will use the terminology ``spin''
when referring to the discrete variables labelling Fourier
components. We will always denote by $w$ (possibly with an index)
the positions of domains of the type described above, and by $z$
(again possibly with an index) the positions of indicator events.
We do not assume a priori that these objects are holomorphic
functions of $w_1,\ldots,w_k$, but, as for the variables
$z_1,\ldots,z_l$, we will omit the dependence on
$\b{w}_1,\ldots,\b{w}_k$ for notational convenience. We also
define \beq
    Q^{(0,l)}(z_1,\ldots,z_l) \,=\, P(z_1,\ldots,z_l) \;,\qquad\qquad Q^{(0,0)}\,=\,1\label{Qzero}~.
\eeq

We then have the following theorem, which is our main result:
\begin{theorem} \label{theoWI}
Let $P(K\cap E_{w_1,\ep_1,\theta_1}\neq \emptyset,\ldots,K\cap
E_{w_k,\ep_k,\theta_k}\neq\emptyset, z_1,\ldots,z_l)$ denote a
probability of intersection of events in a conformal restriction
measure with exponent $h$ on connected subsets $K\in\uH$,
$\{0,\infty\}\in\b{K}$, with $E_{w,\ep,\theta}$ subsets of $\uH$
as defined in (\ref{defE}), (\ref{g}), and $z_1,\ldots,z_l$
representing $l$ indicator events. With the assumptions
\ref{assump}, we have that the limit (\ref{Qmultiple}) for
$n_1=n_2=\ldots=n_k=2$ exists for all $k\ge0$ and $l\ge 0$, and
that it satisfies the following recursion relations:
\beqa\label{WImultiple}
    && Q^{(k+1,l)}_{2,\ldots,2}(w_1,\ldots,w_{k+1},z_1,\ldots,z_l) \,=\, \n && \qquad\qquad \lt[
        \sum_{i=1}^k \lt(\frc1{w_{k+1}-w_i} - \frc1{w_{k+1}}\rt)\frc{\p}{\p w_i}
    + \sum_{i=1}^k \frc2{(w_{k+1}-w_i)^2} + \rt. \\ &&\qquad\qquad \lt.
        + \sum_{i=1}^l \lt(\frc{1}{w_{k+1}-z_i} -\frc1{w_{k+1}}\rt) \frc{\p}{\p z_i}
        + \sum_{i=1}^l \lt(\frc{1}{w_{k+1}-\b{z}_i} -\frc1{w_{k+1}}\rt) \frc{\p}{\p \b{z}_i}
        + \frc{h}{w_{k+1}^2} \rt] \;\cdot \n && \qquad\qquad\qquad\qquad\qquad
           \cdot\; Q^{(k,l)}_{2,\ldots,2}(w_1,\ldots,w_k,z_1,\ldots,z_l)
   \no
\eeqa for all $k\ge0$ and $l\ge0$. In particular,
$Q^{(k,l)}_{2,\ldots,2}(w_1,\ldots,w_k,z_1,\ldots,z_l)$ are
meromorphic functions of $w_1,\ldots,w_k$.
\end{theorem}

\begin{rema} \em \label{remaConfmaps}
In fact, Theorem \ref{theoWI} could be stated in a still more
general fashion, in two ways.

First, the map (\ref{g}) above, being part of the definition of
the family ${\cal E}$ of regions $E_{w,\ep,\theta}$ considered in
the theorem, can be modified by adding to it any (finite) number
of terms of the type
\[
    A\, \ep^p \lt(\frc{f(\theta)}{(w-z)^q} + \frc{(f(\theta))^*}{(\b{w}-z)^q}
    - \frc{f(\theta)}{w^q} - \frc{(f(\theta))^*}{\b{w}^q}\rt)
\]
for any $A\in\C,~2<p\in \R,~ q\in \N$ with $p\ge q+1$ and any
function $f$ finite on $[0,2\pi]$. If $D$ is chosen to be a disk
centered at $w$ of radius strictly larger than $1/4$, the boundary
of $E_{w,\ep,\theta}$ still describes the same ellipse
(\ref{gshape}) as $\ep\to0$, plus, this time, an additional
deformation $O(\ep^{p-q})$. If $p>q+1$ for all terms added, the
additional deformations are sub-leading, but if $p=q+1$ for some
terms, this may not be true anymore. Given the freedom in the
choice of the initial domain $D$, it is not clear for us to which
extent more freedom is provided by such terms, but it will be
clear that our proof below is not affected by the presence of
these terms.

Second, we could have replaced the scaling map $S_{w,\ep}$ in
(\ref{defE}) by the scaling map $S_{w,\ep^r}$ for any $r>0$,
without affecting the proof of Theorem \ref{theoWI}. Then, with
such a scaling map, we could have added terms to the conformal map
$g$ (\ref{g}) as above, but with the condition $p\ge q+1$ replaced
by $p\ge r(q+1)$ (this is a weaker condition for $r<1$). This
provides much more freedom, and generically the circumference of
the boundary of the associated domains $E_{w,\ep,\theta}$ will
then be proportional to $\ep^r$ as $\ep\to0$. We will not go into
further analysis of this possibility.
\end{rema}

\section{Motivations from SLE}\label{sectmotivation}\setcounter{equation}{0}

The aim of this section is to illustrate the ideas which lead to
the identification of the stress-energy tensor within the SLE
language. The arguments presented here are not rigorous, but have
the advantage of applying to other kinds of operators in CFT as
well. More complete and rigorous arguments for the identification
of the stress-energy tensor will be given in the next sections
through the tool of conformal restriction.

Differently to the rest of the paper, where we consider
probabilities of \textit{intersecting} some domains included in
$\mathbb{H}$, here we will examine the event of \textit{passing
between} the ending points of a segment. The reason is that the
SLE equation for the corresponding probability can be easily
obtained for any value of $\kappa$, even in cases where conformal
restriction does not hold. We will discuss at the end of the
section which are the analogies and differences between the two
cases.

Let us consider a chordal ${\rm SLE}_\kappa$ process (for
$0<\kappa<8$) on the upper half plane $\uH$ described by complex
coordinates $w,\bar{w}$. Let us also consider the
probability\footnote{Notice that we use the calligraphic style
(${\cal P}$ here, and ${\cal Q}$ below) when referring to events
fully characterized by two points, in order to make clear their
distinction from events of intersecting some domain.} ${\cal
P}(w_1,w_2,\bar{w}_1,\bar{w}_2)$ of any event that can be fully
characterized by two points $w_1,\,w_2$ on the upper half plane,
in the sense that it is characterized, after any conformal
transformation $G$, by the two points $G(w_1),\,G(w_2)$. From
Ito's formula, ${\cal P}(w_1,w_2,\bar{w}_1,\bar{w}_2)$ satisfies
the equation
\begin{equation}\label{eq12}
\left\{\frac{\kappa}{2}\left(\partial_{w_1}+\partial_{\bar{w}_1}+\partial_{w_2}+\partial_{\bar{w}_2}\right)^2
+\frac{2}{w_1}\partial_{w_1}+\frac{2}{\bar{w}_1}\partial_{\bar{w}_1}+\frac{2}{w_2}\partial_{w_2}
+\frac{2}{\bar{w}_2}\partial_{\bar{w}_2}\right\}{\cal
P}(w_1,w_2,\bar{w}_1,\bar{w}_2)\,=\,0\;.
\end{equation}
It will be more convenient to parameterize the event by the middle
point $w$ of a straight segment, by its length $\ep$ and by the
angle $\theta$ that it makes with the positive imaginary
direction:
$$
w_1=w-\frac{\epsilon}{2}e^{i\theta}\;,\qquad
w_2=w+\frac{\epsilon}{2}e^{i\theta}\;.
$$
We will now analyze the leading contributions to the expectation
${\cal P}(w,\bar{w}, \epsilon,\theta)$ of such an event as
$\ep\to0$. Assuming that each of the Fourier modes of the
probability, parameterized by the \lq\lq spin" $n$ and defined
as\footnote{We use the notation $\tilde{\ }$ to indicate that we
keep the full $\epsilon$--dependence of the Fourier components,
contrary to taking the limit for $\epsilon\to 0$ as in
(\ref{Qmultiple}).}
\begin{equation}\label{Qnpoints}
\tilde{{\cal
Q}}_n(w,\bar{w},\epsilon)\,\equiv\,\int\limits_{0}^{2\pi}\,d\theta\,e^{-in\theta}\,{\cal
P}(w,\bar{w},\epsilon,\theta)\;,
\end{equation}
vanishes with a power law $\epsilon^{x_n}$ as $\ep\to0$, we can
use $\p_\ep = O(\ep^{-1})$ to extract the leading order of
eq.\,(\ref{eq12}):
\begin{equation}\label{eqP}
\left\{\frac{\kappa}{2}\left(\partial_{w}+\partial_{\bar{w}}\right)^2
+\frac{2}{w}\partial_{w} +\frac{2}{\bar{w}}
\partial_{\bar{w}}-\left(\frac{1}{w^2}+\frac{1}{\bar{w}^2}\right)\epsilon\partial_{\epsilon}+
\left(\frac{1}{w^2}-\frac{1}{\bar{w}^2}\right)i\partial_{\theta}+O(\epsilon^2)\right\}{\cal
P}(w,\bar{w}, \epsilon,\theta)\,=\,0\;.
\end{equation}
Performing a Fourier transform diagonalizes the operator
$\p_\theta$, so that the Fourier modes (\ref{Qnpoints}) satisfy,
to leading order,
\begin{equation}\label{eqQn}
\left\{\frac{\kappa}{2}(\partial_{w}+\partial_{\bar{w}})^2+\frac{2}{w}\,\partial_{w}
+\frac{2}{\bar{w}}\,\partial_{\bar{w}}
-\left(\frac{1}{w^2}+\frac{1}{\bar{w}^2}\right)\,\epsilon\,\partial_{\epsilon}
-n\,\left(\frac{1}{w^2}-\frac{1}{\bar{w}^2}\right)\right\}\,\tilde{{\cal
Q}}_n(w,\bar{w}, \epsilon)+\text{corrections}\,=\,0\;,
\end{equation}
where the corrections will be described below. It is easy to check
that
\begin{equation}\label{Qnsol}
\tilde{{\cal Q}}_n(w,\bar{w}, \epsilon)\,=\,c_n
\,\epsilon^{\,x_n}\,w^{\alpha_n}\,\bar{w}^{\beta_n}\,(w-\bar{w})^{\gamma_n}\;,
\end{equation}
with
\begin{equation}\label{xn}
\alpha_n=\frac{\kappa-8}{2\kappa}-\frac{n}{2}\;,\quad
\beta_n=\frac{\kappa-8}{2\kappa}+\frac{n}{2}\;,\quad
\gamma_n=\frac{(8-\kappa)^2-\kappa^2n^2}{8\kappa}\;,\quad
x_n=1-\frac{\kappa}{8}+\frac{\kappa}{8}\,n^2\;,
\end{equation}
satisfies eq.\,(\ref{eqQn}). In Appendix\,\ref{appcalogero}, we
justify this choice of solution for the events that the SLE curve
passes between the two points (that is, to the left of $w_1$ and
to the right of $w_2$, or viceversa). Note that this does not
determine the actual probability corresponding to each of these
events until one can fix the constants $c_n$. As expected, the
lowest scaling exponent is $x_0=2-d_f$, where
$d_f=1+\frac{\kappa}{8}$ is the fractal dimension of SLE.

The function (\ref{Qnsol}, \ref{xn}) gives the correct solution
for the $n$-th Fourier component up to $O(\ep^{x_n})$ only if the
terms neglected in (\ref{eqQn}) contribute to higher order in
$\epsilon$. This is not automatically guaranteed, since the
discarded terms induce a mixing of Fourier components. By
inspecting the structure of equation (\ref{eq12}), it is easy to
see that the corrections to (\ref{eqP}) only contain terms of the
form $\left(\epsilon^2\,e^{\pm 2i\theta}\right)^m$, with $m\geq
1$. As a consequence, (\ref{eqQn}) gets additional contributions
of the form $\epsilon^{2m}\,{\cal D}\tilde{{\cal Q}}_{n-2m}$,
where ${\cal D}$ is some differential operator of order $O(1)$.
Therefore, (\ref{Qnsol}, \ref{xn}) is the actual solution only for
the values of $n$ such that
\begin{equation}\label{condition}
x_n\,<\,2m+x_{n-2m} \qquad \qquad \forall\; m\geq 1\quad
\text{such that}\quad c_{n-2m}\neq 0\;.
\end{equation}

When the relation
\begin{equation}\label{relkappan}
\kappa\,=\,\frac{8}{n+1}
\end{equation}
holds, (\ref{Qnsol}) simplifies to the purely holomorphic function
\begin{equation}\label{Qnsolhol}
\tilde{{\cal Q}}_n(w, \epsilon)\,=\,\text{const}
\,\times\,\left(\frac{\epsilon}{w}\right)^{n}\;,
\end{equation}
where spin and scaling dimension are equal. This suggests a CFT
interpretation of the leading order in $\ep$ of the event in terms
of purely holomorphic fields, whose physical meaning may be
inferred from relation (\ref{relkappan}). For instance, the
holomorphic probability with $n=1$ appears at $\kappa=4$, which is
known to represent the level lines of a free boson, where the
current is a holomorphic field with precisely spin 1. Another
interesting example is given by $n=\frac{1}{2}$ and
$\kappa=\frac{16}{3}$, suggestive of a fermionic field in the
Fortuin-Kasteleyn representation of the Ising model. From the SLE
point of view, the latter value of the spin can naturally occur by
imposing conditions on the winding of the SLE curve around the two
points; this has the effect of increasing the range of $\theta$
beyond which the probability is periodic. Depending on these
conditions, the Fourier modes of the probability ${\cal
P}(w,\bar{w}, \epsilon,\theta)$ may be nonzero only for even
spins, or only for integer spins, or only for half-integer spins,
etc.

The value $n=2$ corresponds to the case of interest in the present
paper. In the next Sections we shall analyse the case $n=2$ and
$\kappa=\frac{8}{3}$, and we shall justify the identification of
$\tilde{{\cal Q}}_2$ with a CFT correlation function involving the
stress-energy tensor.

As already anticipated, however, in the following we will be
interested in the probability $P^{\text{segm}}(w,\bar{w},
\epsilon,\theta)$ of intersecting the small segment (instead of
passing in between its two ending points), and its Fourier
components $\tilde{Q}^{\text{segm}}_n(w,\bar{w},\epsilon)$,
defined as in (\ref{Qnpoints}). At leading order in $\epsilon\to
0$, $P^{\text{segm}}(w,\bar{w}, \epsilon,\theta)$ satisfies
eq.\,(\ref{eqP}) as well. This can be seen by acting on
$P^{\text{segm}}(w,\b{w},\ep,\theta)$ with the Loewner map and
using Ito's formula, together with the transformation property
\beq\label{transP}
    P^{\text{segm}}(w,\b{w},\ep,\theta)
    \mapsto P^{\text{segm}}\left(G(w),\overline{G(w)},\,|\p G(w)|\ep\,,\,\theta + {\rm
    arg}(\p G(w))\right)
\eeq which holds at leading order in $\epsilon$ for any conformal
map $G$, since locally, a conformal transformation is a
combination of a translation, a rotation and a scale
transformation. Obviously, deformations of the segment induced by
the conformal mapping will alter the higher order structure of
eq.\,(\ref{eqP}), but they do not affect the leading order
behaviour of $\tilde{Q}^{\text{segm}}_2$ for $\kappa=8/3$. By
Theorem \ref{theoWI} this is true if the segment is replaced by a
region $E_{w,\ep,\theta}$, which can be chosen to be a very
elongated ellipse, close to a segment of length $\ep$, plus
deformations of order $\ep^2$, as described in the previous
section. In Appendix\,\ref{appdeformation}, we argue that these
deformations do not affect the leading order of
$\t{Q}^{\text{segm}}_2$.

\section{The general result from conformal restriction}\label{sectRestr}\setcounter{equation}{0}

The case $n=2$ of the result discussed in
Section\,\ref{sectmotivation} is particularly interesting, since
the value 2 is the spin of the stress-energy tensor in conformal
field theory. The corresponding value $\kappa=\frac{8}{3}$ is also
peculiar, being the one at which SLE enjoys the property of {\em
conformal restriction}. We will show that this property alone
implies Eqs. (\ref{WImultiple}), which are of the nature of the
conformal Ward identities found in conformal field theory. In
Section\,\ref{sectinterpr} we will use this and other results in
order to relate these objects to certain type of correlation
functions in conformal field theory involving the stress-energy
tensor. For now, we first recall a more general family of measures
satisfying conformal restriction, of which one member is ${\rm
SLE}_{8/3}$ \cite{LSW}.

\subsection{Conformal restriction measures}

Consider a measure $\mu$ on connected subsets $K\subset\uH$ with
$\{0,\infty\}\subset\b{K}$. The measure satisfies conformal
restriction if \beqa
    S\cdot \mu &=& \mu \n
    \mu|_{K \subset \uH\bs D} &=& \Phi_D^{-1}\cdot \mu
    \label{restr}
\eeqa where $S$ is a scale transformation with center at 0,
$D\subset \uH$ is such that $\uH\bs D$ is simply connected and
contains $0$ and $\infty$, and $\Phi_D:\uH\bs D \to\uH$ is a
conformal map which removes $D$ and preserves $0$ and $\infty$. By
normalizing the map $\Phi_D$ such that $\Phi_D(z)\sim z$ as
$z\to\infty$, (\ref{restr}) implies \cite{LSW}
\begin{equation}\label{restrexp}
    P(K \subset \uH\bs D)\,=\,\left[\Phi_D'(0)\right]^{h}
\end{equation}
where $h$ is called the restriction exponent of $K$. In
particular, SLE$_{8/3}$ has been proven to satisfy conformal
restriction with $h=\frac{5}{8}$.

It is important to realize that conformal restriction can be seen
as a combination of conformal invariance and a restriction
property. Indeed, if we use the symbol $\mu_{{\cal H}}$ to
represent measures on connected sets $K\subset {\cal H}$
connecting $0$ to $\infty$, then it is natural to take conformal
invariance to state that $\mu_{\uH\bs D} = \Phi_D^{-1}\cdot
\mu_{\uH}$, and restriction to state that $\mu_{\uH\bs D} =
\mu_\uH\big|_{K\subset \uH\bs D}$. In other words, conformal
invariance and the restriction property can be seen as two
different ways of relating probabilities defined on the domains
$\uH$ and $\uH\bs D$, and the fact that these two ways should lead
to the same result gives a strong constraint on the measure, which
is conformal restriction.

In the following, however, we will need to consider the case when
$\uH\bs D$ is not simply connected. The conformal restriction
property has recently been considered in multiply-connected
domains of the type $\uH\bs D$ \cite{bf,nonsc}. It was verified
that \beq\label{restrnonsc}
    \mu|_{K\subset \uH\bs D} = G^{-1}\cdot \mu|_{K\subset \uH\bs D'}
\eeq if $D\subset \uH$ and $D'\subset \uH$ are related by
$G(\uH\bs D) = \uH\bs D'$ for some conformal transformation $G$
preserving $0$ and $\infty$ (both points also included in $\uH\bs
D$). This can again be viewed as a combination of conformal
invariance and restriction: $\mu_{\uH\bs D} = G^{-1} \cdot
\mu_{\uH\bs D'}$ and $\mu_{\uH\bs D} =
\mu_{\uH}\big|_{K\subset\uH\bs D}$. From the viewpoint of
statistical models, this is very natural since lattice models
certainly admit a description on multiply connected domains. For
instance, the continuum limit of the critical $O(n)$ model at
$n=0$, if it exists, should still satisfy conformal invariance for
conformal transformations relating domains of this type, and
should exhibit the restriction property relating probabilities on
$\uH \bs D$ to conditioned probabilities on $\uH$.

In this case, conformal invariance and the restriction property do
not form two different ways of relating the same pair of domains,
since the image $\uH\bs D'$ of $\uH\bs D$ under a conformal
transformation cannot be anymore the whole $\uH$. However, their
combination still provides non-trivial constraints, essentially
because there are more conformal transformations $\uH\bs D\to
\uH\bs D'$ relating domains of this type than there are conformal
transformations preserving $\uH$. From a pragmatic point of view,
one can {\em define by restriction} probabilities on $\uH\bs D$
where $D\subset \uH$ and one can {\em verify} that the defined
probabilities are related to each other by conformal invariance.
Note that such a definition of probabilities on multiply-connected
domains would also be possible for any measure, not necessarily
having the conformal restriction property, like SLE$_\kappa$ for
generic $\kappa$. But for $\kappa\neq 8/3$, we would not expect
conformal invariance to hold on the resulting probabilities (for
conformal transformations that do not map $\uH$ to itself).

In much the same way that (\ref{restr}) implies (\ref{restrexp}),
it was shown \cite{nonsc} that (\ref{restrnonsc}) implies
\begin{equation}\label{restrexpnonsc}
    P(K\subset \uH\bs D)\,=\,\left[G'(0)\right]^{h}P(K\subset \uH\bs D')
\end{equation}
where $G:\uH\bs D \to\uH\bs D'$ is such that $G(0)=0$ and
$G(z)\sim z$ as $z\to\infty$.

\subsection{Single slit} \label{ssectSingle}

We now show Theorem \ref{theoWI} in the case $k=0$, under the
assumptions \ref{assump}. The proof requires the use of the
conformal transformation (\ref{g}), which is singular at the
location $w$ of the center of the ellipse. This is natural from
the intuition that the insertion of a stress-energy tensor inside
a correlation function, in conformal field theory, can be seen as
resulting from a (non-globally defined) conformal transformation
that is the identity at infinity and that has a pole at the point
of insertion.

{\em Proof of Theorem \ref{theoWI} in the case $k=0$}. We will
begin by using (\ref{restrexpnonsc}) to calculate $Q_2^{(1,0)}$
and thus prove (\ref{WImultiple}) for both $k=0$ and $l=0$. With
$G = g_{w,\ep,\theta}$ and $D$ replaced by $D_{w,\ep} =
S_{w,\ep}(D)$ as in Sect.\,\ref{sectresult},
Eq.\,(\ref{restrexpnonsc}) reads
$$
    P(K\subset \uH\bs D_{w,\ep})\,=\,
    \left[g_{w,\ep,\theta}'(0)\right]^{h}\,\left\{1-P(w,\ep,\theta)\right\}
$$
where we have introduced the more compact notation
$P(w,\ep,\theta) = P(K\cap E_{w,\ep,\theta}\neq \emptyset)$ (this
probability is not a holomorphic function of $w$, but for
notational convenience, here an below we do not write explicitly
the dependence on $\b{w}$). By applying
$\int\,d\theta\,e^{-2i\theta}$ to both sides of this equation,
using the fact that the left hand side is independent of $\theta$
and expanding in $\epsilon$ (with point 3 of our assumptions
\ref{assump}), we obtain
\begin{equation}\label{Q2exact}
    Q_2^{(1,0)}(w)
    \,=\,\frac{h}{w^2}~.
\end{equation}
Note that this leading behavior as $\ep\to0$ has the same
dependence on $w$ as that of $\t{{\cal Q}}_2(w,\epsilon)$ given by
(\ref{Qnsolhol}) at $\kappa=\frac{8}{3}$ (up to a normalization).
In a similar fashion we obtain, for a generic
$n\in\mathbb{Z}\setminus\{0\}$, that
\begin{equation}\label{Qnexact}
    \int_0^{2\pi} d\theta\,e^{-2in\theta} P(w,\ep,\theta) \,=\, O(\ep^{2|n|}) ~,\quad
    \int_0^{2\pi} d\theta\,e^{-i(2n+1)\theta} P(w,\ep,\theta) \,=\, 0 ~,\quad
\end{equation}
which implies that $Q_{2n}^{(1,0)}(w)$ exists.

Consider again the conformal transformation (\ref{g}). From
invariance of the restricted probabilities under conformal
mappings, we have (we denote $\{z\} = z_1,\ldots,z_l$ and
$P(\cdots)_{\uH\bs D} = P(\cdots | K \subset\uH \bs D)$)
\begin{eqnarray*}
    P(\{z\})_{\uH\bs {D}_{w,\epsilon}}&=& P(g_{w,\epsilon,\theta}(\{z\}))_{\uH \bs E_{w,\epsilon,\theta}}=\\
    &=& P(\{z\})_{\uH \bs E_{w,\epsilon,\theta}}+\frac{\epsilon^2}{16}
    e^{2i\theta}\sum\limits_i\left(\frac{1}{w-z_i}\partial_{z_i}\,+\,
    \frac{1}{w-\bar{z}_i}\partial_{\bar{z}_i}\,-\,\frac{1}{w}(\partial_{z_i}+\partial_{\bar{z}_i})\right)P(\{z\})\\
    &&\,+\,\frac{\epsilon^2}{16}
    e^{-2i\theta}\sum\limits_i\left(\frac{1}{\bar{w}-z_i}\partial_{z_i}+
    \frac{1}{\bar{w}-\bar{z}_i}\partial_{\bar{z}_i}\,-\,\frac{1}{w}(\partial_{z_i}+\partial_{\bar{z}_i})\right)
    P(\{z\}) \,+\,o(\epsilon^2)\;.
\end{eqnarray*}
In order to understand the second step, consider the expression
$$
    \frc{P(g_{w,\epsilon,\theta}(\{z\}))_{\uH \bs E_{w,\epsilon',\theta}} -
    P(\{z\})_{\uH \bs E_{w,\epsilon',\theta}}}{\ep^2}~.
$$
The limit as $\ep\to0$ exists by point 1 of our assumptions
\ref{assump}. Also, the limit as $\ep'\to0$ and the limit as
$\ep\to0$ are independent by point 2. Hence, we can send first
$\ep'\to0$ in order to evaluate the expression using point 3; this
gives the terms with derivatives with respect to $\{z\}$. But we
obtain the same value setting first $\ep'=\ep$ then sending
$\ep\to0$. This explains the second step.

From the definition of restricted probabilities, we can write
$$
    P(\{z\})_{\uH \bs E_{w,\epsilon,\theta}}\,=\,\frac{P(\{z\})-P(\{z\},w,\epsilon,\theta)}{
      1-P(w,\epsilon,\theta)}\;,
$$
where we have introduced the more compact notation
$P(\{z\},w,\ep,\theta) = P(\{z\},K\cap E_{w,\ep,\theta}\neq
\emptyset)$. This implies
\begin{eqnarray}\label{eqPz}
    P(\{z\})_{\uH \bs D_{w,\epsilon}} &=&
    P(\{z\})\,-\,P(\{z\},w,\epsilon,\theta)\,+\,P(w,\epsilon,\theta)\,P(\{z\})\,+\,
    \\\nonumber
    &&\,+\,\sum_{n=1}^{\infty}P(\{z\})P(w,\ep,\theta)^{n+1}-\sum_{n=1}^{\infty}
    P(\{z\},w,\ep,\theta)P(w,\ep,\theta)^{n}+
    \\\nonumber &&\,+\,\frac{\epsilon^2}{16}
    e^{2i\theta}\sum\limits_i\left(\frac{1}{w-z_i}\partial_{z_i}\,+\,
    \frac{1}{w-\bar{z}_i}\partial_{\bar{z}_i}-\frac{1}{w}(\partial_{z_i}+\partial_{\bar{z}_i})\right)\,P(\{z\})+
    \\\nonumber
    &&\,+\,\frac{\epsilon^2}{16}
    e^{-2i\theta}\sum\limits_i\left(\frac{1}{\bar{w}-z_i}\partial_{z_i}+
    \frac{1}{\bar{w}-\bar{z}_i}\partial_{\bar{z}_i}-\frac{1}{\bar{w}}(\partial_{z_i}+\partial_{\bar{z}_i})
    \right)\,P(\{z\})+\\
    \nonumber &&\,+\,o(\epsilon^2) \;.
\end{eqnarray}
Applying to eq.\,(\ref{eqPz}) the integral
$\int\limits_{0}^{2\pi}d\theta\,e^{-2i\theta}$ and using the fact
that the left-hand side is independent of $\theta$, we obtain
\begin{eqnarray}\label{eqQ2extra}
    0&=&\,-\int_0^{2\pi} d\theta\,e^{-2i\theta} \,P(\{z\},w,\ep,\theta)\,
    +\,\frc{\pi}8\,\ep^2\,Q_2^{(1,0)}(w)\,P(\{z\})\,+\,
    \\\nonumber
    && \,+\,\frac{\pi}{8}\,\epsilon^2
    \sum\limits_i\left(\frac{1}{w-z_i}\partial_{z_i}\,+\,
    \frac{1}{w-\bar{z}_i}\partial_{\bar{z}_i}-\frac{1}{w}(\partial_{z_i}+\partial_{\bar{z}_i})\right)\,P(\{z\})\,+\,
    \\\nonumber &&\,+\,o(\epsilon^2)\;
\end{eqnarray}
where we used (\ref{Qnexact}) (for $n=1$) in the first line. We
used the fact that the second line of (\ref{eqPz}) contributes
only to $o(\ep^2)$. In order to see this, consider the first sum
and expand $P(w,\ep,\theta)$ in its Fourier modes (in the variable
$\theta$). Under $\int\limits_{0}^{2\pi}d\theta\,e^{-2i\theta}$,
the terms left are those whose total spin (the sum of the spins of
their factors) is 2. By (\ref{Qnexact}), the leading of these
terms as $\ep\to0$ are those for which all factors have zero spin
except one factor; this gives a contribution $o(\ep^2)$ since
there is at least two factors (and using point 3 of the
assumptions \ref{assump}). Consider now the second sum on the
second line of (\ref{eqPz}). Again using Fourier modes, now the
leading terms will be those for which the total spin of the
Fourier components of $P(w,\ep,\theta)$ is 2, 0 or -2. In the case
2 and -2, using (\ref{Qnexact}) and point 3 of assumptions
(\ref{assump}), the contributions are $o(\ep^2)$. In the case 0,
the contributions are $o(\ep) \cdot \int_0^{2\pi}
d\theta\,e^{-2i\theta} \,P(\{z\},w,\ep,\theta)$ which is of higher
order than the first term in the first line of (\ref{eqQ2extra})
and hence gives contributions to $o(\ep^2)$. Using further the
result (\ref{Q2exact}), we finally obtain
\begin{equation}\label{WIsingle}
    Q_2^{(1,l)}(w,\{z\})\,=\,
    \sum\limits_i\left(\frac{1}{w-z_i}\partial_{z_i}\,+\,
    \frac{1}{w-\bar{z}_i}\partial_{\bar{z}_i}-\frac{1}{w}(\partial_{z_i}+\partial_{\bar{z}_i})
    +\frac{h}{w^2}\right)\,P(\{z\})\;,
\end{equation}
which is the special case $k=0$ of (\ref{WImultiple}). \eproof

\subsection{Multiple slits}\label{ssectMultiple}

In order to prove Theorem \ref{theoWI} for $k\ge 1$, we derive the
way by which the quantity $$Q^{(k,l)}_{2,\ldots,2}(w_1,\ldots
w_k,\{z\}\,|\,K\subset \uH \bs D)~,$$ for some simply connected
$D\subset \uH$ bounded away from $w_1,...,w_k$ with $\partial D$
piecewise smooth\footnote{as assumed for the domains considered in
Section\,\ref{sectresult}.} (with a straightforward extension of
the notation introduced in (\ref{Qmultiple})), transforms under a
conformal transformation that maps $\uH\bs D$ to a subset of
$\uH$. More precisely, we show below the following proposition.
\begin{propo} \label{propotransfoQ2}
The following transformation property holds
 \beq\label{transfoQ2}
    Q^{(k,l)}_{2,\ldots,2}(w_1,\ldots w_k,\{z\}\, |\, K\subset \uH \bs D) =
    \lt(\prod_{i=1}^k [G'(w_i)]^2\rt) Q^{(k,l)}_{2,\ldots,2}(G(w_1),\ldots,G(w_k),\{G(z)\}\,|\,
    K\subset \uH \bs D')
\eeq for $G:\uH\bs D \to \uH \bs D'$.
\end{propo}

Let us first prove Theorem \ref{theoWI} in the general case using
this proposition.

{\em Proof of Theorem \ref{theoWI} in the general case}.
Proposition \ref{propotransfoQ2} is enough to prove
(\ref{WImultiple}) in the general case. Indeed, we just have to
repeat the derivation of equation (\ref{WImultiple}) done in the
previous sub-section in the case $k=0$, but using
$Q^{(k,l)}_2(w_1,\ldots,w_k,\{z\}\,|\,K\subset \uH \bs
D_{w_{k+1},\ep})$ instead of $P(\{z\}\,|\,K\subset \uH \bs
D_{w,\ep})$ as a starting object, and using (\ref{transfoQ2}) with
$G = g_{w,\ep,\theta}$ instead of invariance under the
transformation $g_{w,\ep,\theta}$ as a starting step. The rest of
the derivation goes along similar lines, using our assumptions
\ref{assump} in order to obtain derivatives with respect to
$w_1,\ldots,w_k$ as well as with respect to $z_1,\ldots,z_l$, and
we immediately find \beqa
    && Q^{(k+1,l)}_{2,\ldots,2}(w_1,\ldots,w_k,w_{k+1},z_1,\ldots,z_l) \\ && \qquad = \lt[
        \sum_{i=1}^k \lt(\frc1{w_{k+1}-w_i} - \frc1{w_{k+1}}\rt)\frc{\p}{\p w_i}
        + \sum_{i=1}^k \lt(\frc1{w_{k+1}-\b{w}_i} - \frc1{w_{k+1}}\rt)\frc{\p}{\p \b{w}_i}
    + \sum_{i=1}^k \frc2{(w_{k+1}-w_i)^2} + \rt. \n &&\qquad\qquad \lt.
        + \sum_{i=1}^l \lt(\frc{1}{w_{k+1}-z_i} -\frc1{w_{k+1}}\rt) \frc{\p}{\p z_i}
        + \sum_{i=1}^l \lt(\frc{1}{w_{k+1}-\b{z}_i} -\frc1{w_{k+1}}\rt) \frc{\p}{\p \b{z}_i}
        + \frc{h}{w_{k+1}^2} \rt]\;\cdot \n && \qquad\qquad\qquad\qquad \cdot\;
        Q^{(k,l)}_{2,\ldots,2}(w_1,\ldots,w_k,z_1,\ldots,z_l) ~.
   \no
\eeqa Recursively using the fact that
$Q^{(k,l)}_{2,\ldots,2}(w_1,\ldots,w_k,z_1,\ldots,z_l)$ is
analytic in $w_1,\ldots,w_k$, we obtain (\ref{WImultiple}) and
Theorem \ref{theoWI}. \eproof

\begin{rema}\em \label{remaAltproof}
It is worth mentioning that an alternative proof of the multiple
Ward identity that mimics the proof of the single Ward identity in
sub-Section \ref{ssectSingle} could be obtained along the
following lines. First, find a conformal map with simple poles at
the positions $w_1,\ldots,w_k$ and parameterized by the variables
$\ep_1,\ldots,\ep_k$ and $\theta_1,\ldots,\theta_k$ in such a way
that the domain $\uH \bs (D_1\cup\cdots\cup D_k)$, for some
$D_1,\ldots, D_k$ disjoint simply connected regions of $\uH$, is
mapped into $\uH \bs (E_{w_1,\ep_1,\theta_1}\cup \cdots \cup
E_{w_k,\ep_k,\theta_k})$. Then, apply the techniques of
sub-Section \ref{ssectSingle} by taking the spin-2 Fourier
components for all variables $\theta_1,\ldots,\theta_k$ and by
looking at the leading order when $\ep_1\to0,\ldots,\ep_k\to0$
independently (this should be allowed by the conformal map).
Finally, observe the multiple Ward identity (\ref{WImultiple}) by
comparing what is obtained with $k\mapsto k+1$ and what is
obtained with $k$. In Appendix \ref{app2slits}, we present a part
of the proof along these lines by giving the conformal map that
gives the multiple Ward identity for $k=2$. Unfortunately, we were
as of yet unable to show that this conformal map is able to
produce the region $\uH\bs (E_{w_1,\ep_1,\theta_1}\cup
E_{w_2,\ep_2,\theta_2})$; we believe that for this, one needs to
use the freedom of the choice of conformal maps along the lines of
Remark \ref{remaConfmaps}.
\end{rema}

{\em Proof of Proposition \ref{propotransfoQ2}}. We must first
derive some general properties of maps $f$ from boundaries $\p D$
of disjoint unions of simply connected domains $D=\cup_i D_i
\in\uH$ (such that $\partial D_i$ are piecewise smooth) to the
complex numbers, defined by
\[
    f(\p D) = P(\{z\},K\subset \uH\bs D\,|\,K\subset \uH \bs {\cal D})
\]
for simply connected ${\cal D}\in\uH$ bounded away from 0, from
$\infty$, and from $D$. The first property is as follows. From our
assumptions \ref{assump}, the following limit exists:
\beq\label{lim}
    \lim_{\eta\to0} \frc{f((\id + \eta H)(\p D))-f(\p D)}\eta
\eeq where $H$ is any real-analytic conformal map that maps
$\uH\bs {\cal D}$ to to another domain of $\uH$ of the same
topology (with one hole). In fact, this limit can be written as
appropriate derivatives with respect to the coordinates $x_i$ in
the moduli space of $\uH\bs(\{z\}\cup D \cup {\cal D})$ with $0$
and $\infty$ fixed. There is a finite number of derivatives, and,
choosing appropriate coordinates, every derivative $\p/\p x_i$ can
be obtained by an appropriate small and smooth deformations $\eta
H_i$ of $\p D$. The coefficients of these derivatives are linear
in $H$ (since, first, they are not singular when $H$ is zero
anywhere on $\p D$, and second, one can replace $\eta\mapsto
q\eta$ to see that the result scales linearly with $H$) and they
depend on $H$ only through the image of $\p D$ under $H$. Hence,
they are linear functionals of $H$ supported on $\p D$ and can be
written as integrals on $\p D$ of $H$ times appropriate functions
making the projection onto $H_i(\p D)$. Putting these integrals
together, we can write \beq\label{linearfct}
    \lim_{\eta\to0} \frc{f((\id + \eta H)(\p D))-f(\p D)}\eta = \int_0^1
    ds \,H(\p D(s))\, (\Delta_{s} f)(\p D) + \int_0^1 ds\, H(\overline{\p D}(s))\, (\bar\Delta_{s} f)(\p D)~,
\eeq where $s$ is the normalized length along $\p D$ starting from
any point on $\p D$ and going counter-clockwise on each component
in a fixed order, normalized to a total length of 1, and $\p D(s)$
is the associated value of $\p D$. This equation essentially
defines the new maps $\Delta_{s} f,\, \b\Delta_{s} f$ (for all
$s$).

For the second property that we will need, consider, for $G:\uH\bs
\t{D} \to \uH \bs \t{D}'$ for some $\t{D}\subset D$, \beq
    f(G((\id + \eta H)(\p D)))
    = f(F(G(\p D)))
\eeq where \beq
    F = G \circ (\id + \eta H) \circ G^{-1} = \id + \eta (G' \circ G^{-1})\,(H\circ G^{-1}) +
    O(\eta^2)~.
\eeq We can write \beq
    \lim_{\eta\to 0} \frc{f(F(G(\p D))) - f(G(\p D))}\eta = \int_0^1
    ds \,[G'(\p D(s))]^2 \, H(\p D(s))\, (\Delta_{s} f)(G(\p D)) + c.c.
\eeq where $c.c.$ means ``complex conjugate'' and we used \beq
    [G(\p D)](\t{s}) = G(\p D(s)) \Rightarrow d\t{s} = G'(\p D(s)) ds~.
\eeq Hence, the map $f\circ G$ has the same property as $f$, that
is, \beq
    \lim_{\eta\to0} \frc{(f\circ G)((\id + \eta H)(\p D)) - (f\circ G)(\p D)}\eta =  \int_0^1
    ds \,H(\p D(s))\, (\Delta_{s} (f\circ G))(\p D) + c.c.
\eeq with \beq\label{transfoDelta}
    \Delta_{s} (f\circ G) = [G'(\p D(s))]^2 (\Delta_{s} f)\circ G~.
\eeq

Now, using (\ref{linearfct}) with ${\cal D} = D_{w,\ep}$ and with
$\id + \eta H = g_{w,\ep,\theta}$, $\eta = \ep^2$, we can easily
derive an expression similar to (\ref{WImultiple}) for the
quantity $Q^{(1,l)}_2(w,\{z\},\gamma\subset\uH\bs D)$. In fact, it
is convenient to keep the starting point $a\in\R$ of the curve
arbitrary for now, so that we have \beqa
    && Q^{(1,l)}_2(w,\{z\},\gamma\subset \uH \bs D;a) \n &&
    = \lt[ \sum_i \lt( \frc1{w-z_i}\p_{z_i} + \frc1{w-\b{z}_i} \p_{\b{z}_i} \rt) + \frc1{w-a} (\p_a + \p_{\b{a}})
    + \frc{h}{(w-a)^2} \rt] P(\{z\},\gamma\subset \uH \bs D;a) + \n && \int_0^1
    ds\, \frc1{w-\p D(s)} (\Delta_{s} P)(\{z\},\gamma\subset \uH \bs D;a)
    + \int_0^1
    ds\, \frc1{w-\overline{\p D}(s)} (\bar\Delta_{s} P)(\{z\},\gamma\subset \uH \bs D;a)~.
    \no
\eeqa We can obtain a similar expression for
$Q_2(w,\{G(z)\},\gamma\subset G(\uH \bs D);G(a))$ where $G$ is a
real analytic conformal transformation that maps $\uH \bs D$ to
$\uH \bs D'$ for some $D'\subset \uH$ simply connected, with
$G(z)\sim z$ at $z\to\infty$ (but generically, $G(a)\neq a$):
\beqa
    && Q_2^{(1,l)}(w,\{G(z)\},\gamma\subset G(\uH \bs D);G(a))\n &&
    = \lt[ \sum_i \lt( \frc1{w-G(z_i)}\p_{G(z_i)} +
      \frc1{w-G(\b{z}_i)} \p_{G(\b{z}_i)}\rt) + \frc1{w-G(a)} (\p_{G(a)} + \p_{G(\b{a})})
    + \frc{h}{(w-G(a))^2} \rt] \;\cdot \n &&
    \qquad\qquad \cdot\; P(\{G(z)\},\gamma\subset G(\uH \bs D);G(a)) + \n && \qquad \int_0^1
    ds\, \frc{G'(\p D(s))}{w-G(\p D(s))} (\Delta_{s} P)(\{G(z)\},\gamma\subset G(\uH \bs D);G(a))
    +\n && \qquad \int_0^1
    ds \,\frc{G'(\overline{\p D}(s))}{
      w-G(\overline{\p D}(s))} (\bar\Delta_{s} P)(\{G(z)\},\gamma\subset G(\uH \bs D);G(a))
    \no
\eeqa Using relation (\ref{transfoDelta}), the last two lines can
be written \beqa
    && \int_0^1
    ds \,\frc1{w-G(\p D(s))} \frc1{G'(\p D(s))} (\Delta_{s} (P\circ G))(\{z\},\gamma\subset \uH \bs D;a)
    + \n && \int_0^1
    ds \, \frc1{w-G(\overline{\p D}(s))} \frc1{G'(\overline{\p D}(s))}
    (\bar\Delta_{s} (P\circ G))(\{z\},\gamma\subset \uH \bs D;a)
    \no
\eeqa where $P\circ G$ means the map from $\{z\},\p D,a$ to
$[0,1]$ given by $P(\{G(z)\}, \gamma\subset G(\uH\bs D);G(a))$,
and as before, for the purpose of the symbol $\Delta_s$, it is
regarded as a function of $\p D$. Now consider $G$ such that
$G(a)=a$ so that we can use (\ref{restrexpnonsc}): $P(\{G(z)\},
\gamma\subset G(\uH\bs D);G(a)) = (G'(a))^{-h} P(\{z\},
\gamma\subset \uH\bs D;a)$. Hence, we have \beqa
    && (G'(a))^{h} \, Q^{(1,l)}_2(w,\{G(z)\},\gamma\subset G(\uH \bs D);a)\n &&
    =
    \Bigg[ \sum_i \lt( \frc1{w-G(z_i)}\frc1{G'(z_i)} \p_{z_i} +
      \frc1{w-G(\b{z}_i)} \frc1{G'(\b{z}_i)} \p_{\b{z}_i}\rt) + \n && \qquad \qquad +
        \frc1{w-a} \frc1{G'(a)} (\p_{a} + \p_{\b{a}}) -\frc{h}{w-a} \frc{G''(a)}{(G'(a))^2}
    + \frc{h}{(w-a)^2} \Bigg] \; P(\{z\},\gamma\subset \uH \bs D;a) + \n &&
    \qquad \int_0^1
    ds \, \frc1{w-G(\p D(s))} \frc1{G'(\p D(s))} (\Delta_{s} P)(\{z\},\gamma\subset \uH \bs D;a)
\label{exprtemp}
    + \\ && \qquad \int_0^1
    ds \, \frc1{w-G(\overline{\p D}(s))}
    \frc1{G'(\overline{\p D}(s))} (\bar\Delta_{s} P)(\{z\},\gamma\subset \uH \bs D;a)
    \no
\eeqa

Consider the analytical properties in $w$ of the last expression.
It gives a real-analytic function of $w$ in $G(\C\bs (D \cup
\bar{D})) = \C \bs (E \cup \bar{E})$ with simple poles at
$G(z_i)$'s and $G(\bar{z}_i)$'s, and a double pole at $a$ (one can
check that there is no pole at $\infty$), the residues being
directly read off. For $w \in E$ or $w\in \b{E}$, the expression
gives an analytic function. The difference between the expression
near $\p E$ (at $w=G(\p D(s))$, say) outside of $E$ and the
expression near $\p E$ inside of $E$ is
\[
    2\pi i\, [(G^{-1})'(w)]^2\, [\p D'(s)]^{-1}\,(\Delta_s P)(\{z\},\gamma\subset \uH\bs D;a)
\]
A similar result hold near $\overline{\p E}$. These properties
completely determine the analytical functions of $w$ on both sides
of the cuts at $\p E$ and at $\overline{\p E}$.

Finally, consider the expression \beqa &&
    [(G^{-1})'(w)]^2 \;\cdot \n &&
    \cdot\; \Bigg\{ \lt[ \sum_i \lt( \frc1{G^{-1}(w)-z_i}\p_{z_i} + \frc1{G^{-1}(w)-\b{z}_i} \p_{\b{z}_i}\rt) +
        \frc1{G^{-1}(w)-a} (\p_{a} + \p_{\b{a}})
    + \frc{h}{(G^{-1}(w)-a)^2} \rt]\;\cdot \n && \qquad\qquad \cdot\;
    P(\{z\},\gamma\subset \uH \bs D;a) + \n && \qquad \int_0^1
    ds \,\frc1{G^{-1}(w)-\p D(s)} (\Delta_{s} P)(\{z\},\gamma\subset \uH \bs D;a)
    +\n && \qquad \int_0^1
    ds \, \frc1{G^{-1}(w)-\overline{\p D}(s)} (\bar\Delta_{s} P)(\{z\},\gamma\subset \uH \bs D;a) \Bigg\}~.
    \no
\eeqa It is a simple matter to check that it has the same
singularity and cut structure as (\ref{exprtemp}), hence it is the
same function of $w$. This immediately leads to
\beq\label{transfoQ2g}
    Q^{(1,l)}_2(w,\{z\}, K\subset\uH\bs D) =  [G'(w)]^2\,[G'(0)]^h \,Q_2(G(w),\{G(z)\},K\subset \uH\bs D')~.
\eeq Specializing $D$ to be simply connected, this gives
(\ref{transfoQ2}) in the case $k=1$. Note that (\ref{transfoQ2g})
can also be written \beqa &&
    \lim_{\ep\to0} \frc{8}{\pi\ep^2} \int_0^{2\pi} d\theta\,e^{-2i\theta}\, P(\{G(z)\},K\cap G(E_{w,\ep,\theta})
    \neq \emptyset,K\subset \uH \bs D') \n && \qquad\qquad
    = [G'(w)]^2 \,Q^{(1,l)}_2(G(w),\{G(z)\},K\subset \uH\bs D')~. \no
\eeqa If we take one component of $D$ to be itself some
$E_{\t{w},\t\ep,\t\theta}$ and if we integrate over $\t\theta$
with the factor $8 \ep^{-2} e^{-2i\pi\t\theta}/\pi$, we can use
this same equation to derive \beq
    Q^{(2,l)}_{2,2}(w_1,w_2,\{z\}, K\subset\uH\bs D) =  [G'(w_1)]^2\,
    [G'(w_2)]^2\,[G'(0)]^h \,Q^{(2,l)}_{2,2}(G(w_1),G(w_2),\{G(z)\},K\subset \uH\bs D')~.
\eeq Repeating the process, dividing the left-hand side by
$P(K\subset \uH\bs D)$ and the right-hand side by $P(K\subset
\uH\bs D')$ and using (\ref{restrexpnonsc}), we obtain
(\ref{transfoQ2}) for arbitrary $k$. \eproof

\section{CFT interpretation} \label{sectinterpr}\setcounter{equation}{0}

In this section, we shall interpret Theorem \ref{theoWI} from the
point of view of CFT, showing that it represents the Ward
identities, hence $Q_{2,...,2}^{(k,l)}$ can be identified with
correlation functions involving the stress-energy tensor. Recall
that being based on conformal restriction, Theorem \ref{theoWI}
holds for SLE only at the particular value $\kappa=8/3$, which
corresponds to a CFT with central charge $c=0$, as discussed
below. It is natural that the stress-energy tensor is identified
with a local event in SLE only for $\kappa=8/3$, since this
corresponds to the limit $n\to0$ of the $O(n)$ model, which is the
only limit where the loops disappear and where the domain wall is
sufficient to describe the full CFT. Possible generalizations to
CFT with $c\neq 0$ will be mentioned in Section\,\ref{sectbdry}.

Let us consider in detail the application of Theorem \ref{theoWI}
to SLE$_{8/3}$. The corresponding restriction exponent
$h=\frac{5}{8}$ coincides in CFT with the conformal weight of the
boundary operator $\phi_{2,1}$ at $c=0$, which is the value
associated to $\kappa=\frac{8}{3}$ in the identification
(\ref{c(kappa)}). This particular operator has already been
understood to play an important role in the correspondence between
probabilities in SLE and correlation functions in CFT, being the
one inserted at the points where the SLE curve starts and ends
\cite{BB}.

Therefore, eq.\,(\ref{WIsingle}) (i.e. Theorem \ref{theoWI} for
$k=0$) takes the form of the conformal Ward identity which links
the CFT correlation functions
\begin{eqnarray}
P(\{z\})&=&\frac{\langle\phi_{2,1}(0)\,\phi_{2,1}(\infty)\,\prod\limits_i{\cal
O}_i(z_i)\rangle}
{\langle\phi_{2,1}(0)\,\phi_{2,1}(\infty)\rangle}\;,\\
Q_2^{(1,l)}(w,\{z\})&=&
\frac{\langle\phi_{2,1}(0)\,\phi_{2,1}(\infty)\,\prod\limits_i{\cal
O}_i(z_i)\,T(w)\rangle}
{\langle\phi_{2,1}(0)\,\phi_{2,1}(\infty)\rangle}\;,
\end{eqnarray}
where ${\cal O}_i$ are operators with zero scaling dimension and
$T$ is the bulk stress-energy tensor. Recalling the results of
Section \ref{sectmotivation} (and Appendix \ref{appdeformation}),
this means that the spin-2 Fourier component of the SLE
probability of intersecting a segment of length $\;\epsilon\;$ is
associated to the operator $\;\frac{\pi}{8}\,\ep^2T\;$ as
$\epsilon\to 0$. Similarly, Theorem \ref{theoWI} for $k>0$ has the
form of a multiple Ward identity at $c=0$, where
$Q^{(k,l)}_{2,...,2}(w_1,...,w_k,\{z\})$ is a correlation function
involving $k$ insertions of $T\;$:
\begin{equation}
Q^{(k,l)}_{2,...,2}(w_1,...,w_k,\{z\})\,=\,
\frac{\langle\phi_{2,1}(0)\,\phi_{2,1}(\infty)\,\prod\limits_i{\cal
O}_i(z_i)\,T(w_1)\cdots T(w_k)\rangle}
{\langle\phi_{2,1}(0)\,\phi_{2,1}(\infty)\rangle}
\end{equation}

Notice that the transformation property (\ref{transfoQ2}) itself
identifies $Q^{(k,l)}_{2,...,2}(w_1,...,w_k,\{z\})$ with a
correlation function involving $k$ primary operators with spin 2
and scaling dimension 2, plus $l$ dimensionless primary operators.
In general, the stress energy tensor is not a primary operator,
since an extra term appears in its transformation property (the
so-called Schwarzian derivative). However, this term is
proportional to the central charge $c$, and therefore it
disappears in the present case $c=0$.

A further argument in favor of the above correspondence can be
obtained by generalizing eq.\,(\ref{eqQn}) to multiple segments
and extending it to the class of shapes $E_{w,\epsilon,\theta}$.
The resulting equation
\begin{equation}\label{eqSLEmult}
\left\{
\frac{4}{3}\left(\sum_{i=1}^{k}\partial_{w_i}\right)^2-2\sum_{i=1}^{k}
\left(\frac{2}{w_i^2}-\frac{1}{w_i}\,\partial_{w_i}\right)
\right\}Q^{(k,0)}_{2,...,2}(w_1,...,w_k)\,=\,0
\end{equation}
precisely corresponds to the null-vector equation obtained in CFT
by acting with the appropriate combination of Virasoro
differential operators ${\cal L}_n$ on the correlation function of
interest \cite{BPZ}:
$$
\left(\frac{4}{3}{\cal L}_{-1}^2-2 {\cal
L}_{-2}\right)\,\frac{\langle\phi_{2,1}(0)\,\phi_{2,1}(\infty)\,
T(w_1)\cdots T(w_k)\rangle}
{\langle\phi_{2,1}(0)\,\phi_{2,1}(\infty)\rangle}\,=\,0~.
$$

\section{Boundary stress-energy tensor and Ward identities}\setcounter{equation}{0}
\label{sectbdry}

In this Section, we will briefly review the same problem discussed
in the rest of the paper, but in the simpler situation in which
the slits are connected to the boundary of the domain. In this
case, there is no concept of spin as before, and the segments can
be considered to be vertical. It is now natural to look for some
correspondence between probabilities with scaling behavior
$\epsilon^2$ and the boundary stress-energy tensor, which is an
operator of scaling dimension $2$.

This problem has been already analyzed in \cite{FW}, where the
boundary Ward identities have been proven through conformal
restriction. However, it is worth to study it along the lines of
our previous discussion, in order to get a more general result and
a clearer CFT interpretation.

In \cite{FW}, the Ward identities were obtained by directly
exploiting the conformal map
$$
\Phi(z)\,=\,\sqrt{(z-x)^2+\epsilon^2}-\sqrt{x^2+\epsilon^2}\;,
$$
which removes the vertical segment $[x,x+i \epsilon]$ from the
upper half plane $\mathbb{H}$. Inserted in (\ref{restrexp}), this
map produces the result
\begin{equation}\label{Psinglebdry}
P(x,\epsilon)\,=\,\frac{h}{2}\,\frac{\epsilon^2}{x^2}\,+\,o(\epsilon^2)
\end{equation}
for the probability that a restriction set (with restriction
exponent $h$) intersects a single segment connected to the
boundary.

We will now derive the same result of \cite{FW} in a slightly
different way, which is actually the only one generalizable to the
bulk case. We will exploit another kind of conformal map, similar
to (\ref{g}), which has a pole at the location $x$ of the segment,
i.e. where the stress-energy tensor is inserted in the correlation
functions. Let us therefore introduce the singular conformal
transformation
\begin{equation}\label{gbdry}
g_{x,\epsilon}(z)\,=\,z\,+\,\frac{\epsilon^2}{4}\,\frac{1}{x-z}\;,
\end{equation}
which preserves the boundary and maps the semidisk
$D_{x,\epsilon}$ of radius $\frac{\epsilon}{2}$ around
$x\in\mathbb{R}$ to the vertical segment $[x,x+i \epsilon]$. By
implementing (\ref{gbdry}) and using restriction we obtain
\begin{eqnarray*}
P(\{z\})_{\mathbb{H}\setminus D_{x,\epsilon}}&=&P(\{z\})-P(\{z\},x,\epsilon)+P(\{z\})P(x,\epsilon)+\\
&+&
\frac{\epsilon^2}{4}\sum\limits_i\left(\frac{1}{x-z_i}\partial_{z_i}+
\frac{1}{x-\bar{z}_i}\partial_{\bar{z}_i}-\frac{1}{x}(\partial_{z_i}+\partial_{\bar{z}_i})\right)
P(\{z\})+o(\epsilon^2)\end{eqnarray*} where the notation has the
same meaning as in the bulk case. Since now $\mathbb{H}\setminus
D_{x,\epsilon}$ is simply connected, we can map it to the upper
half plane through the function
$$
\Psi(z)=z-\frac{\epsilon^2}{4}\frac{1}{x-z}\;.
$$
Therefore, conformal restriction also implies
\begin{equation*}
P(\{z\})_{\mathbb{H}\setminus D_{x,\epsilon}}=P(\Psi(\{z\}))=
P(\{z\})-\frac{\epsilon^2}{4}\sum\limits_i\left(\frac{1}{x-z_i}\partial_{z_i}+
\frac{1}{x-\bar{z}_i}\partial_{\bar{z}_i}-\frac{1}{x}(\partial_{z_i}+\partial_{\bar{z}_i})\right)
P(\{z\})+o(\epsilon^2)\;,
\end{equation*}
and this leads to the final relation
\begin{equation}\label{WIsinglebdry}
P(\{z\},x,\epsilon)
\,=\,\frac{\epsilon^2}{2}\left[\sum\limits_i\left(\frac{1}{x-z_i}\partial_{z_i}+
\frac{1}{x-\bar{z}_i}\partial_{\bar{z}_i}-\frac{1}{x}(\partial_{z_i}+\partial_{\bar{z}_i})\right)
+\frac{h}{x^2}\right]P(\{z\})+o(\epsilon^2)\;.
\end{equation}
A result analogous to (\ref{WImultiple}) for the probabilities of
intersecting multiple slits can be obtained along the same lines
discussed for the bulk case:
\begin{eqnarray}\label{WImultiplebdry}
&& P(\{z\},x_1,\epsilon_1,...,x_{k+1},\epsilon_{k+1})
\,=\,\\\nonumber
&&\hspace{1cm}\frac{\epsilon_{k+1}^2}{2}\left[\sum\limits_{i=1}^{k}\left(\frac{1}{x_{k+1}-x_i}
-\frac{1}{x_{k+1}}\right)\partial_{x_i}
+\sum\limits_{i=1}^{k}\frac{2}{(x_{k+1}-x_i)^2}\right.+\\\nonumber
&&\hspace{1.8cm}\left.+\,\sum\limits_{i=1}^{l}\left(\frac{1}{x_{k+1}-z_i}\partial_{z_i}+
\frac{1}{x_{k+1}-\bar{z}_i}\partial_{\bar{z}_i}-\frac{1}{x_{k+1}}(\partial_{z_i}+\partial_{\bar{z}_i})\right)
+\frac{h}{x_{k+1}^2}\right]\,\cdot\,\\\nonumber
&&\hspace{3.5cm}\cdot\,P(\{z\},x_1,\epsilon_1,...,x_{k},\epsilon_{k})\,+\,o(\epsilon_{k+1}^2)\;.
\end{eqnarray}
(restriction also implies that
$P(\{z\},x_1,\epsilon_1,...,x_{k},\epsilon_{k})\,=\,O(\epsilon_1^2\cdot...\cdot\epsilon_k^2
)$).

The CFT interpretation of this results is similar to the one
presented in Section\,\ref{sectinterpr}: in the case of
SLE$_{8/3}$, (\ref{WIsinglebdry}) and (\ref{WImultiplebdry})
correspond to the Ward identities if we associate the segment of
length $\;\epsilon\;$ to the insertion of $\;\frac{1}{2}\,\ep^2
T\;$.

\subsection{SLE$_{\kappa}$ with $\kappa<\frac{8}{3}$}

Actually, the result obtained in \cite{FW} holds for any
restriction measure, and it was also applied to an explicit random
set $K$, which is constructed by adding \lq Brownian bubbles' to
SLE \cite{LSW}. Although SLE$_{\kappa}$ does not satisfy
restriction for $\kappa\neq \frac{8}{3}$, the resulting set $K$
enjoys this property if the Brownian bubbles are attached to the
SLE curve with an intensity $\lambda$ chosen as
\begin{equation}
\lambda\,=\,-\,\frac{(3\kappa-8)(6-\kappa)}{2\kappa}\;,
\end{equation}
which is the negative of the central charge $c$ in
(\ref{c(kappa)}). Since $\lambda$ must be positive and the SLE
curve must be a simple curve, this construction only works for
$\kappa<\frac{8}{3}$, which correspond to $c<0$.

As a matter of fact, our result as stated in Theorem \ref{theoWI}
cannot be extended to this construction, because Point 3 in
Assumptions \ref{assump} does not hold for the set $K$ described
above. However, if we restrict the analysis to the boundary case
considered in this section, then the procedure is unaffected by
relaxing Point 3, since we know from
(\ref{Psinglebdry}) that the probability $P(x,\epsilon)$ that $K$
intersects a segment connected to the boundary vanishes as
$\epsilon^2$ when $\epsilon\to 0$ (the generalization of this
result to the probability $P(\{z\},x,\epsilon)$ is
straightforward).

The result (\ref{WImultiplebdry}) can be interpreted from the CFT point of view in the
following way. The restriction exponent associated to the set $K$
is \cite{LSW}
\beq\label{h}
    h = \frc{6-\kappa}{2\kappa} \;,
\eeq and it coincides with the conformal weight of the boundary
operator $\phi_{2,1}$ at generic $\kappa$. Therefore, the
interpretation of eq.\,(\ref{WIsinglebdry}) goes as for
$\kappa=8/3$. The correspondence is not immediately clear,
however, for the case of multiple slits, because
eq.\,(\ref{WImultiplebdry}) does not display the terms
proportional to the central charge which are now expected since
$c\neq 0$. In particular, from CFT one would expect (\ref{WImultiplebdry}) to be
modified as
\begin{eqnarray}
&& P(\{z\},x_1,\epsilon_1,...,x_{k+1},\epsilon_{k+1})
\,\stackrel{?}= \,\\\nonumber
&&\hspace{1cm}\frac{\epsilon_{k+1}^2}{2}\left[\sum\limits_{i=1}^{k}\left(\frac{1}{x_{k+1}-x_i}
-\frac{1}{x_{k+1}}\right)\partial_{x_i}
+\sum\limits_{i=1}^{k}\frac{2}{(x_{k+1}-x_i)^2}\right.+\\\nonumber
&&\hspace{1.8cm}\left.+\,\sum\limits_{i=1}^{l}\left(\frac{1}{x_{k+1}-z_i}\partial_{z_i}+
\frac{1}{x_{k+1}-\bar{z}_i}\partial_{\bar{z}_i}-\frac{1}{x_{k+1}}(\partial_{z_i}+\partial_{\bar{z}_i})\right)
+\frac{h}{x_{k+1}^2}\right]\,\cdot\,\\\nonumber
&&\hspace{3.5cm}\cdot\,P(\{z\},x_1,\epsilon_1,...,x_{k},\epsilon_{k})\,+\\\nonumber
&&
\hspace{1cm}+\,\frac{\epsilon_{k+1}^2}{2}\,\frac{c}{2}\,\sum\limits_{j=1}^{k}\,\frac{1}{(x_{k+1}-x_{j})^4}\,
\frac{\epsilon_{j}^2}{2}\,P(\{z\},x_1,\epsilon_1,...,\hat{x}_j,\hat{\epsilon}_j,...,x_{k},\epsilon_{k})
\,+\,o(\epsilon_{k+1}^2)\;.
\end{eqnarray}
where $\hat{x_j}$ indicates that the coordinate $x_j$ is missing.

The apparent contradiction is solved by identifying probabilities
with \textit{connected} correlation functions in CFT. Intuitively,
this can be understood by noticing that the set $K$ is connected
itself, therefore probabilities of intersecting regions at large
distance from its starting or ending point vanish instead of
factorizing, a property which is realized by connected
correlations functions in QFT. The same idea is valid also at
$\kappa=\frac{8}{3}$, when $K$ reduces to the SLE curve; in that
case, however, connected correlation functions are equal to
unconnected ones, due to the vanishing of the central charge.

Let us define the connected correlation functions as
\begin{equation}\label{conncf}
\langle T_1\,...\,T_k\,{\cal O}\rangle_{c}\,=\,\langle
T_1\,...\,T_k\,{\cal
O}\rangle\,-\,\sum\limits_{j=2}^{k}\,\sum\limits_{\{\alpha\}\subset
\{1,...,k\} \atop \{\beta\}
=\{1,...,k\}\backslash\{\alpha\}}\langle
T_{\alpha_1}\,...\,T_{\alpha_j}\rangle\,\langle
T_{\beta_1}\,...\,T_{\beta_{k-j}}\,{\cal O}\rangle_{c}\;,
\end{equation}
with $T_i\equiv T(x_i)$ and ${\cal O}=
\frc{\phi_{2,1}(0)\,\phi_{2,1}(\infty)}{\langle
\phi_{2,1}(0)\,\phi_{2,1}(\infty)\rangle}$. It is easy to prove
(see Appendix\,\ref{appconnected}) that, if $\langle
T_1\,...\,T_k\,{\cal O}\rangle$ satisfies the conformal Ward
identities at $c\neq 0$, then $\langle T_1\,...\,T_k\,{\cal
O}\rangle_{c}$ satisfies
\begin{eqnarray}\label{WIn}
&&\langle T_1\,...\,T_k\,T_{k+1}\,{\cal O}\rangle_{c}\,=\,\\
&&=\left\{\sum\limits_{i=1}^{k}\left[\left(\frac{1}{x_{k+1}-x_i}-\frac{1}{x_{k+1}}\right)
\partial_{x_i}+\frac{2}{(x_{k+1}-x_i)^2}\right]+
\frac{h}{x_{k+1}^2}\right\}\,\langle T_1\,...\,T_k\,{\cal
O}\rangle_{c}\;,\nonumber
\end{eqnarray}
which are precisely the Ward identities without anomaly as
obtained from restriction in (\ref{WImultiplebdry}). The same
equation can be obtained for ${\cal O}= \frc{\prod_i
\Or_i(z_i)\,\phi_{2,1}(0)\,\phi_{2,1}(\infty)}{\langle\phi_{2,1}(0)\,\phi_{2,1}(\infty)\rangle}$.
Therefore, it is natural to suggest the identification
\begin{equation}\label{connidentif}
P(\{z\},x_1,\epsilon_1,...,x_k,\epsilon_k)\,=\,\frac{\epsilon_1\cdot...\cdot\epsilon_k}{2^k}\;
\frc{\langle\phi_{2,1}(0)\,\phi_{2,1}(\infty)\,\prod\limits_i{\cal
O}_i(z_i)\,T(x_1)\cdots T(x_k)\rangle_{c}}
{\langle\phi_{2,1}(0)\,\phi_{2,1}(\infty)\rangle}\;.
\end{equation}

However, in order to fully justify the identification
(\ref{connidentif}) we should rule out two other possibilities:
one is that we are actually looking again at a $c=0$ CFT, and the
other is that $P(\{z\},x_1,\epsilon_1,...,x_k,\epsilon_k)$
corresponds to a non-connected correlation function at $c\neq 0$
involving primary spin-2 operators instead of the stress-energy
tensor. This can be done by looking at the analog of the SLE
equation\,(\ref{eqSLEmult}) for multiple slits, when we consider
the random process defined by SLE + loops \cite{FW}: \beqa
\label{SLEloopeq}
 &&   \left\{
\frac{\kappa}{2}\left(\sum_{i=1}^{k}\partial_{x_i}\right)^2-2\sum_{i=1}^{k}
\left(\frac{2}{x_i^2}-\frac{1}{x_i}\,\partial_{x_i} \right)
\right\}P(x_1,\epsilon_1,...,x_k,\epsilon_k)\,+ \\
 &&\qquad\qquad   +\lambda \sum\limits_{j=1}^{k}\,\sum\limits_{\{\alpha\}\subset
\{1,...,k\} \atop \{\beta\} =\{1,...,k\}\backslash\{\alpha\}}
\,\frac{\epsilon_{\alpha_1}\cdot...\cdot \epsilon_{\alpha_j}}{2^j}
\,{\cal T}_j(x_{\alpha_1},...,x_{\alpha_j})\,
    P(x_{\beta_1},\epsilon_{\beta_1},...,x_{\beta_{k-j}},\epsilon_{\beta_{k-j}})
      = 0 \nonumber
\eeqa where \beq
    {\cal T}_j(x_1,...,x_j) =      \,
\,\sum\limits_{s\in\sigma_{j}}\;\frac{1}{x_{s(1)}^{2}(x_{s(2)}-x_{s(1)})^2...
(x_{s(j)}-x_{s(j-1)})^2x_{s(j)}^{2}}\,, \label{Tloops}\eeq with
$\sigma_{j}$ indicating the permutations of $j$ numbers. The
meaning of eq.\,(\ref{SLEloopeq}) can be understood by noticing
that (\ref{Tloops}) is the probability that a Brownian bubble
intersects $j$ of the $k$ slits.

It can be easily checked that (\ref{SLEloopeq}) coincides with the
CFT null-vector equation
$$
\left(\frac{\kappa}{2}{\cal L}_{-1}^2-2 {\cal
L}_{-2}\right)\,\langle T_1\,...\,T_k\,{\cal O}\rangle_{c}\,=\,0~.
$$
in a CFT with central charge $c=-\lambda$ (the proof is presented
in Appendix\,\ref{appconnected}). The need for connected
correlation functions can be understood as follows: the
null-vector equation for $\langle T_1\,...\,T_k\,{\cal O}\rangle$
only reproduces the terms in (\ref{SLEloopeq}) with $j=1$, i.e. it
only takes into account the cases when the Brownian bubble
intersects a single slit. The additional terms in (\ref{conncf})
precisely generate the events of the Brownian bubble intersecting
more slits.

This corroborates (\ref{connidentif}) and the identification of
the density $\lambda$ as the negative of the central charge.

It is now worth to comment possible extension of this result to
the bulk case discussed in the rest of the paper. As we already
mentioned, Theorem \ref{theoWI} cannot be directly applied to the
SLE + bubbles construction, since the corresponding measure does
not satisfy point 3 in Assumptions \ref{assump}. A natural
interpretation of this fact is that the measure on Brownian
bubbles, although it satisfies conformal restriction, does not
exhibit anymore the "Markov property" as SLE; that is, we cannot
partially restrict the random set and say that the rest is
obtained by conformal transformation from the initial domain.
Therefore, a description in terms of a local field theory as for
SLE$_{8/3}$ does not seem possible anymore. At this point, it may
seem puzzling that things work for the boundary case, as shown in
\cite{FW} and further elaborated in this section. However, one
should notice that if a connected set intersects a segment of
height $\epsilon$ connected to the boundary, then the outer
boundary of the connected set necessarily intersects the segment
as well. We think that there exists a correct description giving
bulk connected correlation functions at $c<0$ starting from the outer
boundary of the SLE+Brownian bubbles.

\section{Conclusions}\label{conclusions}\setcounter{equation}{0}

In this paper, we have shown that suitable probabilities in SLE
and related processes can be associated to certain correlation
functions containing the holomorphic stress-energy tensor $T(w)$
of CFT with central charge $c= 0$. Our result can be conceptually
stated as 1) the identification between a particular random
variable and the stress-energy tensor:
$$
\frc{8}{\pi}\,\lim_{\ep\to0} \ep^{-2} \int_0^{2\pi} d\theta\,
e^{-2i\theta}\,v(w,\ep,\theta) \;\leftrightarrow\; T(w)\;,
$$
where $v(w,\ep,\theta)$ is 1 when the random set intersects a
segment centered at $w$ of length $\ep$ and of angle $\theta$ with
respect to the imaginary direction, and 0 otherwise; and 2) the
identification between the stochastic average of such random
variables (in the random processes considered) and correlation
fiunctions in CFT with $c=0$. This result adds to previous ones in
the understanding of the connection between SLE and CFT: the
boundary stress-energy tensor was already identified in \cite{FW},
and the end-points of the SLE curve where first identified with
$\phi_{2,1}$ boundary operators of CFT in \cite{BB}. It can be
generalized in three main directions.

One is the application of our methods to other conformally
invariant processes on the plane or on other Riemann surfaces,
like self-avoiding loops and the Conformal Loop Ensemble (CLE),
whose formalization based on conformal restriction is at present
an active research topic in the mathematical community. In
particular, this should give access to CFT with $c>0$, therefore
to a rigorous derivation of (a wide range of models of) CFT in
terms of stochastic processes. Moreover, an appropriate
generalization of the SLE + Brownian bubbles construction is the
natural candidate for the description of CFT with $c<0$, as we
have seen for the boundary case.

The second natural extension of the present work is the
identification of other kinds of holomorphic operators, which, as
we have seen, naturally emerge at some values of $\kappa$. To
justify their correspondence with local CFT operators one should
prove appropriate functional relations analogous to the Ward
identities derived here.

Finally, another possible direction is the identification of other
primary scaling operators. These can be specified, for example, by
requiring that the SLE curve pass between two given points,
separated by a distance $\epsilon$, in a prescribed manner.
Correlation functions with insertions of these operators will
correspond to the coefficients of given powers of $\epsilon$ in
the expansion of the associated probability as the points approach
each other. One would like to show that the local operators
generated in this way then form a closed operator algebra, and
compute the OPE coefficients directly. This would lead to a
construction of at least one sector of the full CFT from the
viewpoint of conformally invariant measures on planar sets.

\vspace{0.5cm}

\textbf{Acknowledgments}

We thank A. Lef\`{e}vre and W. Werner for useful discussions, and
the anonymous referee for important comments. B.D. would like to
thank E. Dell'Aquila as well for discussions during the Durham
symposium on Geometry, Conformal Field Theory and String Theory,
July-August 2005, and Universit\'e Paris VI (Orsay) for support
during his visit, September 2005. B.D. and V.R. are also grateful
to MAPMO (Universit\'e d'Orl\'eans) for financial support to
attend the workshop ``SLE, percolation and stochastic forms'',
October 2005. This work was supported by EPSRC, under the grants
GR/R83712/01 (V.R. and J.C.) and GR/S91086/01 (B.D., post-doctoral
fellowship).

\begin{appendix}

\section{SLE probabilities in the disk geometry}\label{appcalogero}\setcounter{equation}{0}

The \textit{ansatz}
\begin{equation}\label{Qnsolapp}
\tilde{{\cal Q}}_n(w,\bar{w}, \epsilon)\,=\,c_n
\,\epsilon^{\,x_n}\,w^{\alpha_n}\,\bar{w}^{\beta_n}\,(w-\bar{w})^{\gamma_n}\;
\end{equation}
solves eq.\,(\ref{eqQn}) for two different choices of the
parameters:
\begin{equation}\label{xndiscarded}
\alpha_n=-\frac{2n}{\kappa-4}\;,\quad
\beta_n=\frac{2n}{\kappa-4}\;,\quad \gamma_n=-\frac{2\kappa
n^2}{(\kappa-4)^2}\;,\quad x_n=\frac{2\kappa n^2}{(\kappa-4)^2}
\end{equation}
and
\begin{equation}\label{xnapp}
\alpha_n=\frac{\kappa-8}{2\kappa}-\frac{n}{2}\;,\quad
\beta_n=\frac{\kappa-8}{2\kappa}+\frac{n}{2}\;,\quad
\gamma_n=\frac{(8-\kappa)^2-\kappa^2n^2}{8\kappa}\;,\quad
x_n=1-\frac{\kappa}{8}+\frac{\kappa}{8}\,n^2\;.
\end{equation}
In order to select the correct set of parameters, it is convenient
to map our problem onto the unit disk $\uD$, through the
transformation $z' = \frc{z-w}{z-\b{w}}\,$ for $z\in\uH\,$ and
$z'\in\uD$. This transformation maps the point $w$ to the center
of the disk, the length $\ep$ to $\ep/|w-\b{w}|$, and it shifts
the angle $\theta$ by an angle of $\pi/2$. Also, the point $0$ is
mapped to $w/\b{w}$ on the boundary of the disk, and the point
$\infty$ to $1$. We are then describing an SLE curve on the unit
disk started at $w/\b{w}$ and required to end at 1. Fixing the
power of $\ep/|w-\b{w}|$ to be some number $x_n$ (the ``scaling
dimension''), we are left, after integration over $\theta$ as in
(\ref{Qnpoints}), with a second order ordinary differential
equation in the angle $\alpha=\arg(w/\b{w})\in [0,2\pi]$. This
equation is the eigenvalue equation for an eigenfunction of the
two-particle Calogero-Sutherland Hamiltonian with eigenvalue
(energy) $2 x_n/\kappa$ and with total momentum $n$ \cite{johnCS}.
For generic $\kappa$, the Calogero-Sutherland Hamiltonian admits
only two types of series expansions $C\alpha^\omega[[\alpha^2]]$
(with $C\neq0$) as $\alpha\to0^+$ for its eigenfunctions: one with
a leading power $\omega=8/\kappa-1$, the other with a leading
power $\omega=0$. It admits the same two types of series
expansions $C'(2\pi-\alpha)^{\omega'}[[(2\pi-\alpha)^2]]$ (with
$C'\neq0$) as $\alpha\to2\pi^-$. Allowing only one type of series
expansion at $0$ and only one at $2\pi$ (the possibilities give
the Calogero-Sutherland system in the fermionic sector
$\omega=\omega'=8/\kappa-1$, bosonic sector $\omega=\omega'=0$ or
mixed sector, $\omega\neq\omega'$), the Calogero-Sutherland
Hamiltonian has a discrete set of eigenfunctions, with eigenvalues
bounded from below (since it is a self-adjoint operator on the
space of functions with these asymptotic conditions). The lowest
eigenvalue is obtained for the eigenfunction (the ground state)
with the least number of nodes (zeros of the eigenfunction). If
the leading powers $\omega$ and $\omega'$ are chosen equal to each
other, then the ground state (in the sector with total momentum
$n$) is described by the solutions (\ref{Qnsolapp}) with
(\ref{xndiscarded}) (for $\omega=0$) or (\ref{xnapp}) (for
$\omega=8/\kappa-1$), which, in the coordinates of the disk, take
the form
\begin{equation}\label{Qndisk}
\tilde{{\cal Q}}_n(|w-\bar{w}|,\alpha, \epsilon)\,=\,\tilde{c}_n
\,\left(\frac{\epsilon}{|w-\bar{w}|}\right)^{\,x_n}\,e^{i\,\frac{\alpha_n-\beta_n}{2}\,\alpha}
\,\left(\sin\frac{\alpha}{2}\right)^{\gamma_n+x_n}\;.
\end{equation}
The probabilities that we are considering require the curve to
pass by the center of the disk. Hence, they vanish when the
starting point of the SLE curve is brought toward its ending point
on the disk, from any direction; this fixes the power to be
$8/\kappa-1$ (for $\kappa<8$) at both values $\alpha=0,2\pi$ and
therefore selects the solution in the fermionic sector
(\ref{xnapp}). Note that since the probability could be given by
an excited state in the fermionic sector (which corresponds to a
higher value in place of the exponent $x_n$), we {\em do not} have
the condition that $\t{c}_n$ is nonzero.

\section{Deformation of the segment}\label{appdeformation}\setcounter{equation}{0}

In this Appendix, we will show that Theorem \ref{theoWI} can be
used to conclude that the second Fourier component of the
probability $P^{\text{segm}}(w,\bar{w},\epsilon,\theta)$ that the
SLE$_{8/3}$ curve \textit{intersects} a segment is given by
\begin{equation}\label{Q2segm}
\tilde{Q}_2^{\text{segm}}(w,\bar{w},\epsilon)\,=\,\frac{\pi}{8}\,\epsilon^2\,\frac{h}{w^2}\,+\,o(\epsilon^2)\;.
\end{equation}
This means in particular that (\ref{Q2segm}) is equal, at leading
order in $\epsilon$, to the second Fourier component of the
probability ${\cal P}(w,\bar{w},\epsilon,\theta)$ of
\textit{passing between the ending points of the segment} as in
(\ref{Qnsolhol}) with $n=2$, up to an overall constant.

First, let us recall that the result (\ref{WImultiple}), and in
particular (\ref{Q2exact}), applies to the case when the
considered shapes are deformed segments, which correspond to $b\to
1$ in (\ref{gshape}):
\begin{eqnarray}\label{segmdef}
    g_{w,\ep,\theta}\left(w+\frac{\epsilon}{4}\,e^{i\alpha+i\theta}\right) &=&
    w+\frac{\epsilon}{2}e^{i\theta+i\pi/2}\,\sin\alpha
    \,+\,\\\nonumber
    &-&\frac{\epsilon^2}{16}\,\frac{e^{2i\theta}}{w}
    \,-\,\frac{\epsilon^2}{16}\,\frac{e^{-2i\theta}}{\bar{w}}\,+\,
        \frac{\epsilon^2}{16}\,\frac{e^{-2i\theta}}{\bar{w}-w}
    \,+\,\frac{\epsilon^3}{64}\,\frac{e^{i\alpha+i\theta}}{(\bar{w}-w)^2}\,+\,O(\epsilon^4)\;,
\end{eqnarray}
where $\alpha\in[0,2\pi]$. As we have discussed in the main text,
the probability of intersecting a straight segment, corresponding
to the first line of (\ref{segmdef}), satisfies at leading order
eq.\,(\ref{eqP}), and its Fourier components satisfy
eq.\,(\ref{eqQn}), which coincide with the equations for the
probability of passing in between the two ending points of the
segment. We now have to show that the deformations described in
the second line of (\ref{segmdef}) do not affect the leading order
behaviour in (\ref{Q2segm}).

Let us first analyze the effect of the $\epsilon^2$ terms in
(\ref{segmdef}). Since they do not depend on $\alpha$, they merely
correspond to a change in the central position of the segment:
$$
w\,\to\,w\,-\,\frac{\epsilon^2}{16}\,\frac{e^{2i\theta}}{w}
    \,-\,\frac{\epsilon^2}{16}\,\frac{e^{-2i\theta}}{\bar{w}}\,+\,
        \frac{\epsilon^2}{16}\,\frac{e^{-2i\theta}}{\bar{w}-w}\;.
$$
Therefore, their effect on the differential equation (\ref{eqP})
for $P^{\text{segm}}(w,\bar{w},\epsilon,\theta)$ translates into
the introduction of terms of the type
$\epsilon^2\,\partial_w\,\tilde{Q}^{\text{segm}}_m(w,\bar{w},\epsilon)$
and
$\epsilon^2\,\partial_{\bar{w}}\,\tilde{Q}^{\text{segm}}_m(w,\bar{w},\epsilon)$
in the equation (\ref{eqQn}) for
$\tilde{Q}^{\text{segm}}_2(w,\bar{w},\epsilon)$. Since each
Fourier component is assumed to vanish with a power law as
$\epsilon\to 0$, these corrections turn out to be of order
$o(\epsilon^2)$.

The remaining terms in (\ref{segmdef}), of order $\epsilon^3$ and
higher, depend on $\alpha$, therefore they induce a change not
only in the position of the segment, but also in its length and
inclination. However, these can only introduce in (\ref{eqQn})
contributions of the form
$(\epsilon^2+o(\epsilon^2))\,\epsilon\partial_{\epsilon}\tilde{Q}^{\text{segm}}_m(w,\bar{w},\epsilon)$,
$(\epsilon^3+o(\epsilon^3))\,\partial_w\tilde{Q}^{\text{segm}}_m(w,\bar{w},\epsilon)$,
$(\epsilon^3+o(\epsilon^3))\,\partial_{\bar{w}}\tilde{Q}^{\text{segm}}_m(w,\bar{w},\epsilon)$
and
$(\epsilon^3+o(\epsilon^3))\,m\,\tilde{Q}^{\text{segm}}_m(w,\bar{w},\epsilon)$,
which give corrections of order $o(\epsilon^2)$ to
$\tilde{Q}^{\text{segm}}_2(w,\bar{w},\epsilon)$. Furthermore, the
segment gets distorted by these terms in (\ref{segmdef}), so that
it develops higher moments besides the dipole one. However,
assuming smoothness of the probabilities, these contributions are
also of order $o(\epsilon^2)$.

\section{The double Ward identities}\label{app2slits}\setcounter{equation}{0}

In this appendix, we will sketch a possible proof of the multiple
Ward identities (\ref{WImultiple}) alternative to the one
presented in Section \ref{ssectMultiple}, as mentioned in
Remark\,\ref{remaAltproof}. The discussion is not rigorous, but it
displays interesting features that is worth to comment. For
simplicity, we will just consider the case of two slits, but the
following arguments can be easily extended to $k$ slits.

The basic idea is to consider the generalization $g\equiv
g_{w_1,\epsilon_1,\theta_1,w_2,\epsilon_2,\theta_2}$ of the
conformal map (\ref{g}) which is singular at the two points $w_1$
and $w_2$ and satisfies \beq\label{defE2}
    g\left[\uH\bs\left( D_{w_1,\ep_1}\cup D_{w_2,\ep_2}\right)\right] = \uH \bs
    \left(E_{w_1,\ep_1,\theta_1}\cup
    E_{w_2,\ep_2,\theta_2}\right)\;,
\eeq where the notation is the same as in
Section\,\ref{sectresult}, $D_{w_i,\epsilon_i}=S_{w_i,\ep_i}(D_i)$
and $E_{w_i,\epsilon_i,\theta_i}\in {\cal E}$. We can now slightly
extend (\ref{restrexpnonsc}) to write
\begin{eqnarray}\label{restrexpgen}
&& P(\{z\},K\subset\uH \bs
    \left(E_{w_1,\ep_1,\theta_1}\cup
    E_{w_2,\ep_2,\theta_2}\right))\,=\,\\\nonumber&&\hspace{2cm}=\,[(g^{-1})'(0)]^{h}
    \,P(\{g^{-1}(z)\},K\subset\mathbb{H}\backslash\left(
D_{w_1,\ep_1}\cup D_{w_2,\ep_2}\right))\;.
\end{eqnarray}
Since
\begin{eqnarray}\nonumber
P(\{z\},K\subset\uH \bs
    \left(E_{w_1,\ep_1,\theta_1}\cup
    E_{w_2,\ep_2,\theta_2}\right))&=&1\,-\,P(\{z\},K\cap E_{w_1,\epsilon_1,\theta_1}\neq \emptyset)\,-\,
    P(\{z\},K\cap E_{w_2,\epsilon_2,\theta_2}\neq \emptyset)
    \,+\,\\
    &&+\,P(\{z\},K\cap E_{w_1,\epsilon_1,\theta_1}\neq \emptyset,K\cap E_{w_2,\epsilon_2,\theta_2}\neq
    \emptyset)\;,
\end{eqnarray}
eq.\,(\ref{restrexpgen}) implies
\begin{eqnarray}\label{Q22}
    Q^{(2,l)}_{2,2}(w_1,w_2,\{z\}) &=&
    \lt(\frc{8}\pi\rt)^2\,\lim_{\ep_1,\ep_2\to0}
    \ep_1^{-2} \ep_2^{-2} \;
    \int_0^{2\pi} d\theta_1\,e^{-2i\theta_1}
    \int_0^{2\pi} d\theta_2\,e^{-2i\theta_2}\,\cdot\\\nonumber
   &&\cdot [(g^{-1})'(0)]^{h}
    \,P(\{g^{-1}(z)\},K\subset\mathbb{H}\backslash\left(
D_{w_1,\ep_1}\cup D_{w_2,\ep_2}\right))~.
\end{eqnarray}
Therefore $Q_{2,2}^{(2,l)}$ will be expressed as a differential
operator acting on $P(\{z\})$, and the operator is obtained by
expanding the map $g^{-1}$ in $\epsilon_1$ and $\epsilon_2$.

The lack of rigor in our considerations is due to the fact that,
although we know that the map $g$ exists, we do not know its
explicit form. However, we can approximate it with another
conformal map $\hat{g}$, associated to a family of shapes
$\hat{{\cal E}}$ and defined through its inverse as
\begin{eqnarray}
\hat{g}^{-1}(z)&=&
z\,-\,\frac{\epsilon_1^2\,e^{2i\theta_1}}{16}\,\left(\frac{1}{w_1-z}-\frac{1}{w_1}\right)
\,-\,\frac{\epsilon_1^2\,e^{-2i\theta_1}}{16}\,\left(\frac{1}{\bar{w}_1-z}-\frac{1}{\bar{w}_1}\right)
\,+\\\nonumber &&
-\,\frac{\epsilon_2^2\,e^{2i\theta_2}}{16}\,\left(\frac{1}{w_2-z}-\frac{1}{w_2}\right)
\,-\,\frac{\epsilon_2^2\,e^{-2i\theta_2}}{16}\,\left(\frac{1}{\bar{w}_2-z}-\frac{1}{\bar{w}_2}\right)
\,+\\\nonumber
&&\,+\,\frac{\epsilon_1^2\,\epsilon_2^2\,e^{2i\theta_1}\,e^{2i\theta_2}}{(16)^2}\,
\frac{1}{(w_1-w_2)^2}\,\left(\frac{1}{w_1-z}+\frac{1}{w_2-z}-\frac{1}{w_1}-\frac{1}{w_2}\right)\,+\,\\\nonumber
&&\,+\,\frac{\epsilon_1^2\,\epsilon_2^2\,e^{-2i\theta_1}\,e^{-2i\theta_2}}{(16)^2}\,
\frac{1}{(\bar{w}_1-\bar{w}_2)^2}\,\left(\frac{1}{\bar{w}_1-z}+\frac{1}{\bar{w}_2-z}-
\frac{1}{\bar{w}_1}-\frac{1}{\bar{w}_2}\right)\;.
\end{eqnarray}
Let us define the domains $\h{E}_1$ and $\h{E}_2$ as two disjoint
simply connected domains such that \beq
    \hat{g}\left[\uH\bs\left( D_{w_1,\ep_1}\cup D_{w_2,\ep_2}\right)\right] = \uH \bs
    \left(\h{E}_1\cup
    \h{E}_2\right)
\eeq ($\h{E}_1$ and $\h{E}_2$ are disjoint for $\ep_1$ and $\ep_2$
small enough). Both domains $\h{E}_1$ and $\h{E}_2$ depend on the
variables $w_1,\ep_1,\theta_1,w_2,\ep_2,\theta_2$ (as well, of
course, as on the initial domains $D_1$ and $D_2$). It can be
easily checked, however, that $\h{E}_1$ is given at leading order
by $E_{w_1,\epsilon_1,\theta_1}$, plus higher order corrections
which also depend on $w_2$, $\epsilon_2$ and $\theta_2$ (and that
the converse is true for $\hat{E}_2$). If we assume that the
$\theta_2$-dependence of $P(\{z\},K\cap \h{E}_1\neq \emptyset)$
and that the $\theta_1$-dependence of $P(\{z\},K\cap \h{E}_2\neq
\emptyset)$ contribute to the double integration in $\theta_1$ and
$\theta_2$ at higher order in $\epsilon_1\,\epsilon_2$ than
$P(\{z\},K\cap \hat{E}_1\neq \emptyset,K\cap \hat{E}_2\neq
\emptyset)$, we can still use (\ref{Q22}) to obtain
\begin{equation}\label{WIdouble}
\hat{Q}_{2,2}^{(2,l)}(w_1,w_2,\{z\})\,=\,({\cal D}_1+{\cal
D}_2+{\cal D}_3+{\cal D}_4)\,P(\{z\})\;,
\end{equation}
where
\begin{eqnarray*}
    &&\hat{Q}^{(2,l)}_{2,2}(w_1,w_2,\{z\}) = \n &&
    \lt(\frc{8}\pi\rt)^2\,\lim_{\ep_1,\ep_2\to0}
    \ep_1^{-2} \ep_2^{-2} \;
    \int_0^{2\pi} d\theta_1\,e^{-2i\theta_1}
    \int_0^{2\pi} d\theta_2\,e^{-2i\theta_2}\,
    P(\{z\},K\cap \hat{E}_1\neq \emptyset,K\cap \hat{E}_2\neq\emptyset)~,
\end{eqnarray*}
\begin{eqnarray*}
{\cal D}_1&=& \sum_{ij}\,
\left(\frac{1}{w_1-z_i}-\frac{1}{w_1}\right)\left(\frac{1}{w_2-z_j}-\frac{1}{w_2}\right)\partial_i\,\partial_j\,+\,
\sum_{ij}\,\left(\frac{1}{w_1-\bar{z}_i}-\frac{1}{w_1}\right)\left(\frac{1}{w_2-z_j}-\frac{1}{w_2}\right)
\bar{\partial}_i\,\partial_j\,+\,\\
&+&\,\sum_{ij}\,\left(\frac{1}{w_1-z_i}-\frac{1}{w_1}\right)\left(\frac{1}{w_2-\bar{z}_j}-\frac{1}{w_2}\right)
\partial_i\,\bar{\partial}_j\,+\,
\sum_{ij}\,\left(\frac{1}{w_1-\bar{z}_i}-\frac{1}{w_1}\right)\left(\frac{1}{w_2-\bar{z}_j}-\frac{1}{w_2}\right)
\bar{\partial}_i\,\bar{\partial}_j\;,
\end{eqnarray*}
$$
{\cal
D}_2\,=\,\frac{1}{(w_1-w_2)^2}\,\sum_i\,\left[\left(\frac{1}{w_1-z_i}+\frac{1}{w_2-z_i}
-\frac{1}{w_1}-\frac{1}{w_2}\right)
\partial_i+\left(\frac{1}{w_1-\bar{z}_i}+\frac{1}{w_2-\bar{z}_i}-\frac{1}{w_1}
-\frac{1}{w_2}\right)\bar{\partial}_i\right]\;,
$$
\begin{eqnarray*}
{\cal
D}_3&=&\frac{h}{w_1^2}\,\sum_i\,\left[\left(\frac{1}{w_2-z_i}-\frac{1}{w_2}\right)\partial_i
+\left(\frac{1}{w_2-\bar{z}_i}-\frac{1}{w_2}\right)\bar{\partial}_i\right]\,+\\
&+&\frac{h}{w_2^2}\,\sum_i\,\left[\left(\frac{1}{w_1-z_i}-\frac{1}{w_1}\right)\partial_i
+\left(\frac{1}{w_1-\bar{z}_i}-\frac{1}{w_1}\right)\bar{\partial}_i\right]\;,
\end{eqnarray*}
and
$$
{\cal D}_4\,=\,\frac{h^2}{w_1^2 w_2^2}\,+\,\frac{2h}{w_1 w_2
(w_1-w_2)^2}\;.
$$
The result (\ref{WIdouble}) is the symmetrized form of the Ward
identities (\ref{WImultiple}) for $k=1$ with the right-hand side
expanded using the Ward identity (\ref{WImultiple}) for $k=0$. To
make this arguments a proof of (\ref{WImultiple}) one would need
to rigorously justify the assumption before (\ref{WIdouble}) and
to show that $\hat{Q}_{2,2}^{(2,l)}$ coincides with
$Q_{2,2}^{(2,l)}$. In order to do this, it could be useful to
exploit the freedom in the choice of conformal maps as commented
in Remark\,\ref{remaConfmaps}.

\section{Properties of connected correlation functions in CFT}\label{appconnected}\setcounter{equation}{0}

In this Appendix we will explicitly prove that appropriate
connected correlation functions in CFT, defined in (\ref{conncf}),
satisfy equations (\ref{WIn}) and (\ref{SLEloopeq}) presented in
the main text.

Let us first notice that solving the recursion in definition
(\ref{conncf}) we obtain
\begin{eqnarray*}
    \bra T_1 \cdots T_k \Or\ket_c &=&
    \sum_{n=0}^\infty (-1)^n \sum_{\cup_{i=0}^n J_i =
    \{1,\ldots,k\}\atop J_i\cap J_j = \varnothing (i\neq j)}
    \bra T_{J_0}\Or\ket \bra T_{J_1}\ket\cdots
    \bra T_{J_n}\ket \,=\,\\
    &=&
    \sum_{n=0}^\infty \frc{(-1)^n}{n+1} \sum_{\cup_{i=0}^n J_i =
    \{0,\ldots,k\}\atop J_i\cap J_j = \varnothing (i\neq j)}
    \bra T_{J_0}\ket \bra T_{J_1}\ket\cdots
    \bra T_{J_n}\ket
\end{eqnarray*}
where $T_{J_i}\equiv T_{\alpha_1}\cdots T_{\alpha_{|J_i|}}$ with
ordered $\alpha_l\in J_i$ and, in the last equation, $T_0 = \Or$
by definition. Note that the last equation is completely
symmetric: nothing makes the operator $\Or$ particular with
respect to the $T$'s, so that we could as well have correlation
functions connected to any of these $T$'s. In the SLE context,
$\Or$ stands for
$$
\Or\,=\,\frac{\phi_{2,1}(0)\phi_{2,1}(\infty)}{\bra\phi_{2,1}(0)\phi_{2,1}(\infty)\ket}
$$
and we are in the boundary CFT on the half-plane.

We will now show by induction that these connected correlation
functions of energy-momentum tensors in CFT satisfy (\ref{WIn}),
which can be compactly written as
\begin{equation}\label{WardCFTconnected}
    \bra T_1\cdots T_{k+1}\Or\ket_c = {\cal L}_{-2}(x_{k+1}) \bra T_1\cdots
    T_{k}\Or\ket_c~,
\end{equation}
where we have defined the operator
\begin{equation}
    {\cal L}_{-2}(x) \,=\,\sum\limits_{i=1}^{k}\left[\left(\frac{1}{x-x_i}-\frac{1}{x}\right)\partial_i
    +\frac{2}{(x-x_i)^2}\right]+\frac{h}{x^2}\;.
     \label{L-oper}
\end{equation}
Assume that the insertion of the operator $T_l$ in $\bra T_1
\cdots T_l \Or\ket_c$ is implemented by applying the operator
${\cal L}_{-2}(x_l)$ as in (\ref{L-oper}) on the correlation
function $\bra T_1 \cdots T_{l-1} \Or\ket_c$ for all $l\leq k$.
From CFT, we know that
\begin{equation}\label{WardCFT}
    \bra T_1\cdots T_{k+1}\Or\ket = {\cal L}_{-2}(x_{k+1}) \bra T_1\cdots
    T_{k}\Or\ket + \sum_{j=1}^{k} \bra T_j T_{k+1}\ket \bra
    T_1\cdots \widehat{T_j} \cdots T_{k}\Or\ket\;,
\end{equation}
where the symbol $\widehat{T_j} $ means that the operator $T_j$
has been removed from the correlation function. Applying ${\cal
L}_{-2}(x_{k+1})$ on $\bra T_1\cdots T_k\Or\ket_c$, using
(\ref{WardCFT}) and noticing that the inductive hypothesis implies
\begin{eqnarray*}
    {\cal L}_{-2}(x_{k+1}) \bra T_{\alpha_1}\cdots T_{\alpha_j}\ket \bra T_{\beta_1}\cdots
    T_{\beta_{k-j}}\Or\ket_c &=&
    \bra T_{\alpha_1}\cdots T_{\alpha_j}\ket \bra T_{\beta_1}\cdots
    T_{\beta_{k-j}}T_{k+1}\Or\ket_c + \\
    &&
    +\bra T_{\alpha_1}\cdots T_{\alpha_j}T_{k+1}\ket \bra T_{\beta_1}\cdots
    T_{\beta_{k-j}}\Or\ket_c - \\
    &&
    -\sum_{l=1}^j \bra T_{\alpha_k} T_{n+1}\ket
    \bra T_{\alpha_1}\cdots \widehat{T_{\alpha_l}}\cdots T_{\alpha_j}\ket
    \bra T_{\beta_1}\cdots T_{\beta_{k-j}}\Or\ket_c
\end{eqnarray*}
we indeed find (\ref{WardCFTconnected}). It is easy to check
explicitly that this formula is valid for $k=1$, hence the
induction is complete.

\vspace{0.5cm}

In order to prove that $\langle T_1\,...\,T_{j}\,{\cal
O}\rangle_{c}$ also satisfies eq.\,(\ref{SLEloopeq}), we have to
preliminary identify the CFT correlation function corresponding to
${\cal T}_j(x_1,...,x_j)$. By adapting the inductive argument
presented above to the case ${\cal O}=T(0)$, it is straightforward
to check that
\begin{eqnarray*}
&&\langle T_1\,...\,T_{j}\,T_{j+1}\,T(0)\rangle_{c}\,=\,\\
&&=\left\{\sum\limits_{i=1}^{j}\left[\left(\frac{1}{x_{j+1}-x_i}-
\frac{1}{x_{j+1}}\right)\partial_i+
\frac{2}{(x_{j+1}-x_i)^2}\right]+\frac{2}{x_{j+1}^2}\right\}\,\langle
T_1\,...\,T_{j}\,T(0)\rangle_{c}\;.
\end{eqnarray*}
Since $ \langle T(x) T(0)\rangle\,=\,\frac{c/2}{x^4}$, the only
solution to the recursion is
$$
\langle T_1\,...\,T_{j}\,T(0)\rangle_{c}\,=\,\frac{c}{2}\,{\cal
T}_j(x_1,...,x_j)
$$
with ${\cal T}_j$ defined in (\ref{Tloops}).

Therefore, eq.\,(\ref{SLEloopeq}) can be written as
$$
\left\{{\cal
D}-\sum\limits_{i=1}^{k}\frac{4}{x_i^2}\right\}\,\langle
T_1...T_k\,{\cal O}\rangle_c\,+\,
2\,\frac{\lambda}{c}\,\sum\limits_{j=1}^{k}\,\sum\limits_{\{\alpha\}\subset
\{1,...,k\} \atop \{\beta\}
=\{1,...,k\}\backslash\{\alpha\}}\langle
T_{\alpha_1}...T_{\alpha_{j}}\,T(0)\rangle_c \,\langle
T_{\beta_1}...T_{\beta_{k-j}}\,{\cal O}\rangle_c\,=\,0\;,
$$
where we have defined the differential operator
$$
{\cal
D}\,=\,\frac{\kappa}{2}\left(\sum\limits_{i=1}^{k}\partial_i\right)^2+\sum\limits_{i=1}^{k}\frac{2}{w_i}\,\partial_i\;.
$$
We know from CFT that
\begin{equation}\label{nullvectorCFT}
\left\{{\cal
D}-\sum\limits_{i=1}^{k}\frac{4}{x_i^2}\right\}\langle T_1...
T_k{\cal O}\rangle\,-\,
c\,\sum\limits_{i=1}^{k}\,\frac{1}{x_i^4}\,\langle
T_{1}...\widehat{T_{i}}... T_{k}{\cal O}\rangle\,=\,0
\end{equation}
By using the induction hypothesis for $k-j<k$ we have
\begin{eqnarray}\nonumber
&&{\cal D}\,\langle
T_{\beta_1}...T_{\beta_{k-j}}{\cal O}\rangle_c\,=\,\\
&&=\,-2\frac{\lambda}{c}\,\sum\limits_{l=1}^{k-j}\,\sum\limits_{\{\gamma\}\subset
\{\beta_1,...,\beta_{k-j}\} \atop \{\delta\}
=\{\beta_1,...,\beta_{k-j}\}\backslash\{\gamma\}}\langle
T_{\gamma_1}...T_{\gamma_l}\,T(0)\rangle_c\,\langle
T_{\delta_1}...T_{\delta_{k-j-l}}{\cal O}\rangle_c
\,+\,\left(\sum\limits_{l=1}^{k-j}\frac{4}{x^2_{\beta_{l}}}\right)\langle
T_{\beta_1}...T_{\beta_{k-j}}{\cal O}\rangle_c \nonumber\;,
\end{eqnarray}
while CFT tells us that
\begin{eqnarray*}
&&{\cal D}\,\langle
T_{\alpha_1}...T_{\alpha_j}\rangle\,=\,\\
&&=\,-2\,\left[\langle T(0)
T_{\alpha_1}...T_{\alpha_j}\rangle-\sum\limits_{\ell=1}^{j}\langle
T(0)\,T_{\ell}\rangle\langle
T_{\alpha_1}...\widehat{T_{\ell}}\,...T_{\alpha_j}\rangle\right]
\,+\,\left(\sum\limits_{\ell=1}^{j}\frac{4}{x^2_{\alpha_{\ell}}}\right)\langle
T_{\alpha_1}...T_{\alpha_j}\rangle\;.
\end{eqnarray*}
Therefore we have
\begin{eqnarray*}
&&\left\{{\cal
D}-\sum\limits_{i=1}^{k}\frac{4}{x_i^2}\right\}\sum_{\alpha,\beta}\langle
T_{\alpha_1}...T_{\alpha_j}\rangle\,\langle T_{\beta_1}...
T_{\beta_{k-j}}{\cal O}\rangle_c=\\
&-&2\,\sum\limits_{j=2}^{k}\sum\limits_{\alpha,\beta}\langle
T(0)\,T_{\alpha_1}...T_{\alpha_j}\rangle\,\langle
T_{\beta_1}...T_{\beta_{k-j}}{\cal O}\rangle_c+\\
&+&2\,\sum\limits_{j=2}^{k}\sum\limits_{\alpha,\beta}\sum\limits_{\ell=1}^{j}\langle
T(0)\,T_{\ell}\rangle\langle
T_{\alpha_1}...\widehat{T_{\ell}}\,...T_{\alpha_j}\rangle\,\langle
T_{\beta_1}...T_{\beta_{k-j}}{\cal O}\rangle_c-\\
&-&2\frac{\lambda}{c}\,\sum\limits_{j=2}^{k}\sum\limits_{\alpha,\beta}\,\sum\limits_{l=1}^{k-j}
\sum\limits_{\gamma,\delta} \langle
T_{\alpha_1}...T_{\alpha_j}\rangle\,\langle
T_{\gamma_1}...T_{\gamma_l}T(0)\rangle_c\,\langle
T_{\delta_1}...T_{\delta_{k-j-l}} {\cal O}\rangle_c\,
\end{eqnarray*}
and the thesis follows if $\lambda=-c$. The induction is then
completed by checking explicitly (\ref{SLEloopeq}) in the case
$k=2$.

\end{appendix}


\begin{thebibliography}{}


\bibitem{BPZ} Belavin A.A., Polyakov, A.M., Zamolodchikov, A.B.
Nucl. Phys. B \textbf{241} (1984) 333; J. Stat. Phys. \textbf{34}
(1984) 763.

\bibitem{yellow} Di Francesco, P., Mathieu, P., Senechal, D.,
\textit{Conformal Field Theory}, (Springer, Berlin, 1997).

\bibitem{loewner} Loewner, K., Math. Ann. \textbf{89} (1923) 103.

\bibitem{schramm} Schramm, O., Israel J. Math. \textbf{118} (2000) 221.

\bibitem{review} Cardy, J., Annals Phys. \textbf{318} (2005)
81.

\bibitem{JC1984} J. Cardy, Nucl. Phys. B \textbf{240} (1984) 514.

\bibitem{BB} Bauer, M., Bernard, D., Phys. Lett. B \textbf{543} (2002)
135; Commun. Math. Phys. \textbf{239} (2003) 493; Phys. Lett. B
\textbf{557} (2003) 309; Annales Henri Poincare \textbf{5} (2004)
289.

\bibitem{fracdim} Rohde, S., Schramm, O., Annals Math \textbf{161} (2) (2005) 883;

Beffara, V., Annals Prob. \textbf{32} (2004) 2606;

Beffara, V., \textit{math.PR/0211322}.

\bibitem{LSW} Lawler, G., Schramm, O., Werner, W.,
J. Amer. Math. Soc. \textbf{16} (2003) 917.

\bibitem{FW} Friedrich, R., Werner, W., Comm. Math. Phys. \textbf{243} (1) (2003) 105.

\bibitem{bf} Bauer, R., Friedrich, R., \textit{math.PR/0408157}.

\bibitem{nonsc} Beffara, V., \textit{Mouvement brownien plan, SLE, invariance conforme et dimensions fractales},
PhD thesis, available at
\textit{http://www.umpa.ens-lyon.fr/~vbeffara/index.php}.

Lawler, G., \textit{The Laplacian-b random walk and the
Schramm-Loewner evolution}, available at
\textit{http://www.math.cornell.edu/~lawler/}.

\bibitem{johnCS} Cardy, J., Phys. Lett. B \textbf{582} (1-2) (2004) 121.



\end{thebibliography}
\end{document}